\documentclass{article}
\usepackage[utf8]{inputenc}
\usepackage{enumerate} %
\usepackage{arydshln} 
\usepackage{xcolor}
\usepackage{amsmath}
\usepackage{multirow}
\usepackage{amssymb}
\usepackage{authblk} 
\usepackage{lineno}
\usepackage[numbers,sort&compress]{natbib}
\usepackage{graphicx}
\usepackage{doi}
\usepackage{enumerate}
\DeclareUnicodeCharacter{2060}{\nolinebreak}

\usepackage[utf8]{inputenc} 
\usepackage[T1]{fontenc}    
\usepackage{hyperref}       
\hypersetup{backref=true,       
    pagebackref=true,               
    hyperindex=true,                
    colorlinks=true,                
    breaklinks=true,                
    urlcolor= blue,                
    linkcolor= blue,                
    bookmarks=true,                 
    bookmarksopen=false,
    filecolor=black,
    citecolor=blue,
    linkbordercolor=red}
\AtBeginDocument{\hypersetup{pdfborder={0 0 0}}}   
\usepackage{setspace}
\usepackage{geometry}
\title{\textbf{Review of the Higgs boson production via vector boson fusion and its decay into bottom quarks in $pp$ collisions at} \(\boldsymbol{\sqrt{s}=8,13}\) \textbf{TeV}}
\author[$1$]{Fikriye C I Kaya \thanks{corresponding author: \texttt{higgsvbfhbbreview@gmail.com}}}
\author[$2$]{Ibrahim Mirza }

\affil[$1$]{\textit{\small{Department of Mechanical and Electrical Engineering, Oslo Metropolitan University, Oslo 0890, Norway}}}
\affil[$2$]{\textit{\small{Department of Physics and Astronomy, University of Tennessee, Knoxville, Tennessee 37916, USA}}}
\date{}

\begin{document}
\maketitle
\begin{abstract}

We present a comprehensive overview of the vector boson fusion production and the following decay of the Higgs boson to two bottom quarks (\(H\!\to\! b\bar{b}\)), which dominates the Higgs total width yet remains experimentally challenging because of substantial QCD multijet backgrounds. We overview the use of boosted decision trees, which advance jet reconstruction and flavour tagging by including boosted topologies and deep-learning-based \(b\)-tagging and which improve sensitivity in associated production channels.
Special attention is given to ATLAS and CMS selections and results pertaining to the vector boson fusion production of Higgs, decaying in the $b\bar{b}$-channel, from both the 8 TeV and the 13 TeV LHC data-sets conducted with up to 90.8 $fb^{-1}$ of Data and published through 2024.
Consistent with the Standard Model gauge symmetry \(SU(3)_C \times SU(2)_L \times U(1)_Y\) predictions, we emphasise how \(H\!\to\! b\bar{b}\) measurements contribute, in conjunction with other production modes and decay channels, to tests of the bottom-quark Yukawa coupling, which is an essential probe of the electroweak symmetry-breaking mechanism.
Finally, we discuss how the current precision constrains deviations in the bottom-quark Yukawa coupling, outline ongoing efforts during Run 3, and consider the prospects of forthcoming high-luminosity LHC operation and the proposed Future Circular Collider in further tightening precision tests of the Standard Model and beyond.

\end{abstract}
 
\section{Introduction} 



The \emph{Standard Model (SM)} of particle physics is a renormalisable quantum
field theory based on the gauge group
\(\mathrm{SU(3)}_{\mathrm C}\times\mathrm{SU(2)}_{\mathrm L}\times\mathrm{U(1)}_{\mathrm Y}\)%
~\cite{glashow61,weinberg67,salam68,glashow70}.
Electroweak symmetry breaking is instigated by introducing a complex scalar
doublet, established by the Higgs field, whose vacuum expectation value
breaks the symmetry to \(\mathrm{SU(3)}_{\mathrm C}\times\mathrm{U(1)}_{\text{EM}}\)%
~\cite{englert64,guralink64,higgs66}.
The quantum excitation of this field is a neutral, spin-zero scalar boson, namely
the Higgs boson.  Fermion masses arise via Yukawa interactions between the
Higgs field and the fermions.

At the \emph{Large Hadron Collider (LHC)} the Higgs boson is produced through
several mechanisms:
\emph{gluon–gluon fusion (ggF)};
\emph{vector boson fusion (VBF)};
\emph{associated production with a vector boson (VH: \(WH\) or \(ZH\))}, where $V$ denotes the electroweak vector boson and $H$ denotes the Higgs; \emph{production with top/bottom-quark pairs} \(\bigl(t\bar{t}H,\;b\bar{b}H\bigr)\);
and \emph{production with a single top quark} \(\bigl(tH\bigr)\)~\cite{ATLAS:2024fkg}.

The 2012 discovery of the Higgs boson relied on clean final states,
notably \(H\to ZZ^{*}\to4\ell\), where the final state has four \emph{leptons} ($l$) and \(H\to\gamma\gamma\), where the final state has two \emph{photons} ($\gamma$) \cite{CMS:2012qbp,ATLAS:2012yve}.
Its dominant decay,
\(H\to b\bar{b}\), accounts for about \(58\,\%\) of the total width at
\(m_{H}\approx125\,\text{GeV}\)%
~\cite{ParticleDataGroup:2024cfk,PDG2024},
but is experimentally challenging because of large \emph{quantum chromodynamics (QCD)}
multijet backgrounds.
Observation of \(H\to b\bar{b}\) was first achieved in 2018
via \emph{VH} production, where the leptonic decay of the vector boson
enabled efficient triggering and background suppression%
~\cite{ATLAS:2018kot,CMS:2018nsn}. The most recent findings regarding the characteristics of the Higgs-boson are summarised in Table~\ref{tab:1}.

Precise measurement of the bottom-quark Yukawa coupling is critical for the confirmation of the SM. Although the third-generation Yukawa interactions (with $t$, $b$, and $\tau$) have all been directly established, the \emph{fractional} uncertainty on the bottom-quark Yukawa coupling $y_b$ (equivalently the coupling modifier $\kappa_b$) remains the largest among $\kappa_t,\kappa_b,\kappa_\tau$ and is therefore a primary target for improved precision in Run~3 and at the \emph{High-Luminosity LHC (HL-LHC)}~\cite{ATLAS:2022vkf,HiggsComb:Run2,CMS:Run2Comb}.

Recent advances in jet reconstruction and flavour tagging have improved
the sensitivity to \(H\to b\bar{b}\).
Boosted topologies, large-radius jets, and subjet \(b\)-tagging,
combined with machine-learning discriminants
(e.g.\ \emph{boosted decision trees (BDTs)}, neural-network taggers),
enhance the rejection of QCD backgrounds%
~\cite{ATLAS:2019bwq}.
Results are increasingly reported within the
\emph{Simplified Template Cross-Section (STXS)} framework,
which partitions phase space into fiducial regions that minimise
theoretical uncertainties%
~\cite{Berger:2019wnu}.

This paper focuses on the \(H\to b\bar{b}\) decay, which has the largest single contribution to the total Higgs width,  produced via the colour-singlet \emph{VBF} process, characterised by two forward, high-\(p_{T}\) quark-initiated jets and suppressed hadronic activity in the central region. As \emph{VBF} is the second-largest Higgs production mode at the LHC, measurements are based on Data collected up to and including LHC Run 2 and the most recent publications through 2024 \cite{ATLAS:2016mzy,ATLAS:2018jvf,CMS:2015two,CMS:2023tfj}.

\section{ATLAS $\&$ CMS Experiments} 


The \emph{Higgs-to-bottom-quark ($H\to b\bar b$)} programme at the
\emph{Large Hadron Collider (LHC)} relies on two multi-purpose
detectors, the \emph{A Toroidal LHC Apparatus (ATLAS)} and
\emph{The Compact Muon Solenoid (CMS)}.
Both experiments are optimised for heavy-flavour identification,
background suppression, and wide kinematic coverage.

Each detector employs a multi‑layer tracking system around the interaction point.
ATLAS determines the primary vertex with topological algorithms; in the 13 TeV \emph{VBF} analysis the \emph{Jet Vertex Tagger (JVT)} is also used to suppress jets from pile‑up interactions \cite{ATLAS:2016mzy,ATLAS:2018jvf}. 
CMS reconstructs all interaction vertices in each event and then selects as primary the one whose associated tracks maximise the scalar sum of squared transverse momenta, $\sum p_{\mathrm{T}}^{2}$; the same criterion is quoted in both the 8 TeV and 13 TeV analyses \cite{CMS:2015two,CMS:2023tfj}. Tracks are bent in solenoidal fields (2–3 \% momentum resolution in
ATLAS, 1–2 \% in CMS), and refined vertex algorithms provide efficient secondary-vertex reconstruction, which is vital for $b$-tagging \cite{ATLAS:2024ytk,CMS:2019Phase1Tracking,ATLAS:2020IDAlign}.

Offline $b$-jet identification combines several low-level taggers.
ATLAS establishes a \emph{secondary vertex (SV)} b-tagging algorithm, namely \emph{SV1}, and the \emph{JetFitter} algorithm for a topological reconstruction of jets. Impact parameters from these are then fed into either of the two high-level tagging algorithms: The \emph{multivariate (MV)} algorithm \emph{MV2} employs a \emph{boosted-decision-tree (BDT)}, and the \emph{deep learning (DL)} algorithm \emph{DL1\textsubscript{r}} utilizes a deep feed-forward \emph{neural network} \cite{ATLAS:2020ixf,Hartman:2891112,ATLAS:2019btag}.
CMS uses the \emph{Inclusive Vertex Finder (IVF)} algorithm together with two \emph{deep neural network (DNN)} classifiers for b-tagging: \emph{DeepCSV}, which utilizes a \emph{combined secondary vertex (CSV)}, and 
\emph{DeepJet} \cite{CMS:2017wtu,Bols:2020bkb,CMS-DP-2018-058}. The legacy triggers still operate a \emph{likelihood-based} CSV algorithm, where classical multivariate taggers also remain in use for discriminating jets \cite{cMS:2015ebl,CMS:2023tfj}. Complementary to CSV, the CMS 8 TeV also employ the \emph{track counting high-efficiency (TCHE)} for the \emph{High-Level Trigger (HLT)} on \emph{calorimeter-based jets (CaloJets)} as well as on \emph{particle flow (PF)} jets \cite{CMS:2015two}.

Calorimetry in both experiments provides up to $4\pi$ coverage:
ATLAS employs \emph{Lead (Pb)} with  \emph{liquid Argon (LAr)} modules for the electromagnetic section and steel–scintillator tiles (barrel region) or
copper/tungsten absorbers (end-cap/forward regions) for hadron measurements
\cite{ATLAS:1999uwa}.
CMS uses lead-tungstate crystals for the electromagnetic calorimeter
and brass/scintillator (barrel region) or steel/quartz-fiber (end-cap/forward regions) sections for the hadronic calorimeter \cite{CMS:2006myw}.
These systems, combined with anti-$k_{t}$ jet reconstruction
($R=0.4$ for all jets in the \emph{VBF} selections, $R=0.8$–1.0 for separate boosted topologies, where the \emph{radius parameter (R)} quantifies the maximum angular extent of detected jets),
deliver energy measurements up to an \emph{absolute pseudorapidity} ($|\eta|$) of $\sim 5$
\cite{ATLAS:2017ywy,CMS:2017wtu,Bishara:2016kjn}.
In ATLAS Run 2 configuration, including the \emph{Insertable B‑layer (IBL)} pixel detector that was installed in 2014, the all‑silicon pixel–strip tracker covers $|\eta|<2.5$~\cite{LaRosa2016JINST,ATLAS:2016mzy,ATLAS:2018jvf}.
The LAr electromagnetic calorimeter reaches $|\eta|<3.2$,
the TileCal hadronic system extends to $|\eta|<1.7$, and the forward LAr modules in the \emph{Forward Calorimeter (FCal)} bring hadronic coverage out to
$|\eta|\simeq4.9$ \cite{ATLAS:2023ikq}.
CMS offers a comparable reach \cite{CMS:2023tfj}:
Its Phase‑1 upgraded all‑silicon tracker operates within $|\eta|<2.5$. 
A homogeneous PbWO$_{4}$ crystal \emph{electromagnetic calorimeter (ECAL)} and a brass/scintillator sampling \emph{hadron calorimeter (HCAL)} with the barrel and end‑cap regions cover up to $|\eta|<3$, while forward hadron calorimetry in the range $3<|\eta|<5.2$ is provided by the iron/quartz‑fibre \emph{Hadron Forward (HF)} calorimeters \cite{Focardi:2000tr,CMS:2008xjf,CMS:2012qbp}. 

The muon spectrometer reconstructs tracks up to $|\eta|<2.7$ with trigger acceptance $|\eta|<2.4$ \cite{ATLAS:2023ikq}.  CMS muon chambers cover $|\eta|<2.4$ \cite{CMS:2012qbp}. Muon \emph{identification (ID)} is aided by large magnets that are the toroidal magnets in ATLAS and the steel yoke magnets in CMS. Although muons are not part of the $H\to b\bar b$ signal, their ID tightens control of the top and $W/Z$ backgrounds with leptonic decays. ATLAS employs a hardware \emph{Level-1 (L1)} trigger followed by a
software-based HLT \cite{ATLAS:2023gog}, while CMS performs all software selection in its HLT; both systems use jet and \emph{missing transverse momentum ($E_{T}^{miss}$)} signatures, and CMS applies further online criteria \cite{CMSTrigger:2005yhe}.

Reconstruction in ATLAS within the presented analyses follow topological cluster criteria for jets and track thresholds for charged particles \cite{ATLAS:2016mzy}. In the 13 TeV analysis, electrons are reconstructed with a \emph{sliding window} algorithm, whereas the muons are reconstructed after combined calorimetry and muon spectrometry measurements \cite{ATLAS:2018jvf}. Also important to note here: Introduced in 2021 and adopted as the default for Run 3 data-taking in 2022, ATLAS now employs a \emph{unified flow object (UFO)} reconstruction that merges tracking and calorimeter information while retaining topological clustering, which can thus significantly improve analyses \cite{ATLAS:2020gwe}. CMS continues to refine the PF jet-reconstruction, combining signals from all sub-detectors \cite{CMS:2017yfk}.

Both collaborations improve on the $b$-jet energy scale and resolution within calibrations with corrections such as the $b$-jet regression in CMS and the semileptonic $b$-hadron corrections in ATLAS \cite{CMS:2019uxx,ATLAS:2021piz}. Analyses presented here, targeting \emph{VBF}, require at least two $b$-tagged jets consistent with the Higgs mass, accompanied by two well-separated forward jets.

Therefore, through precise vertexing, wide acceptance, advanced triggering and calibrated jet reconstruction, ATLAS and CMS achieve competitive sensitivities to the dominant $H\to b\bar b$ decay in \emph{VBF} (besides \emph{ggF+jets} and \emph{VH}) as detailed in the analyses
reviewed here.

\section{The Vector Boson Fusion Production Mechanism and the Bottom Quark Pair Decay of the Higgs} 


\subsection{Signal Topology and Selection}

In 2012 the ATLAS and CMS experiments at the LHC observed a scalar
particle consistent with the Higgs boson,
confirming the Higgs mechanism \cite{ATLAS:2012yve,CMS:2012qbp}. Although discovery relied on bosonic modes such as
\(H\!\to\! ZZ^{*}\) and \(H\!\to\!\gamma\gamma\), determining the Higgs couplings to quarks remains essential.
With a branching fraction \(\mathcal{B}(H\!\to\! b\bar b)=58.2\pm1.3\%\) at
\(m_{H}\simeq125\,\mathrm{GeV}\),
\(H\!\to\! b\bar b\) has the largest single contribution to the total Higgs width, providing a direct probe of fermion mass generation \cite{PDG2024,ParticleDataGroup:2024cfk}.
Its first observation came in 2018 through
\emph{VH}, with the leptonic decay of the accompanying $V$ \cite{ATLAS:2018kot,CMS:2018nsn}.

Among production modes, \emph{VBF} offers a
distinctive topology: two incoming quarks exchange a \emph{colour-singlet} $V$ in a \(t\)-channel process, yielding two high-\(p_{T}\) forward jets separated by a large rapidity gap and a high dijet invariant mass \cite{CMS:2023tfj}.
The Higgs boson is produced centrally and typically decays to two \(b\)-jets.
While higher-order radiation can alter the topology, large \emph{invariant mass of the two jets} (\(m_{jj}\)) and \emph{pseudorapidity separation} (\(\Delta\eta_{jj}\)) remain powerful signatures.

\begin{figure}[p]
    \centering
    \includegraphics[width=0.5\linewidth]{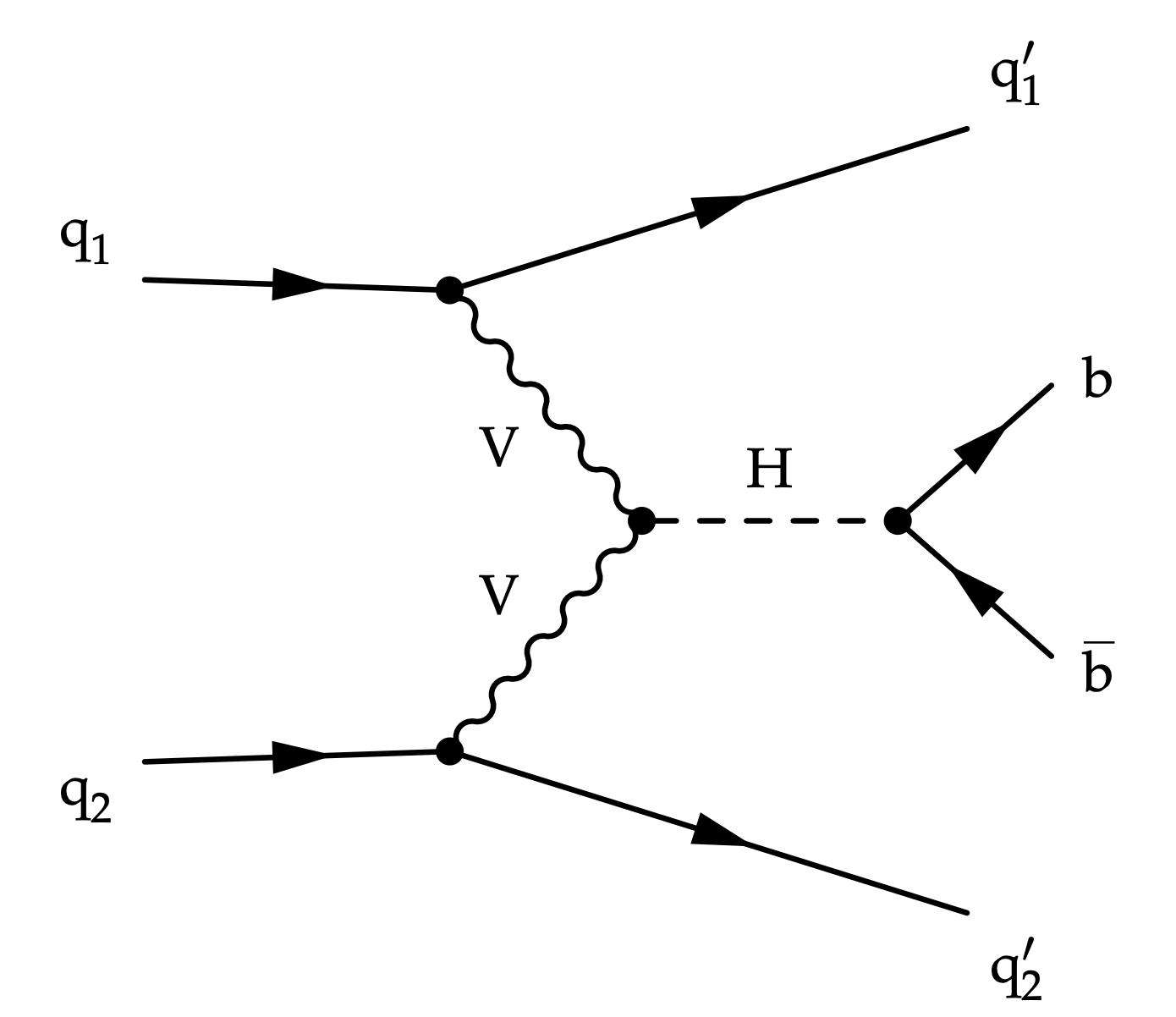}
     \caption{Feynman diagram representation of a \emph{VBF} production of Higgs in Leading Order (LO), followed by a b-flavour quark-pair decay of the Higgs boson ($H\rightarrow b\bar{b}$) \cite{CMS:2023tfj}.}
    \label{fig:one}
\end{figure}

Figure~\ref{fig:one} shows the defining kinematic features of the \emph{VBF} Higgs production, followed by a decay in the $b\bar{b}$-channel, where the subsequent \(H\!\to\! b\bar{b}\) decay proceeds through the bottom-quark Yukawa coupling. Two energetic forward jets with a large rapidity gap point to a colour–singlet electroweak exchange, while the Higgs boson is produced centrally and decays to a pair of \(b\)-flavoured jets.

The ATLAS and CMS analyses discussed here seek \emph{VBF}-like behaviours as follows. Some of the analyses select \emph{Tight} samples, which contain events passing through more stringent trigger and offline selections than the \emph{Loose}, vs. \emph{Loose} samples, which fail the \emph{Tight} cuts but still make it to the event sample (See \cite{ATLAS:2016mzy,ATLAS:2018jvf,CMS:2015two,CMS:2023tfj} for specific selections). The \emph{VBF} selections require two energetic forward jets with a large invariant mass
and rapidity gap.  Across the four analyses the offline thresholds vary:
\begin{itemize}
  \item ATLAS 8 TeV: no explicit $m_{jj}$ cut offline, but the HLT requires $m_{jj}>650\;\text{GeV}$ \cite{ATLAS:2016mzy}.  
  
  \item ATLAS 13 TeV: \emph{photon}$+$\emph{VBF} channel imposes $m_{jj}>800\;\text{GeV}$, while the \emph{two‑} and \emph{four‑central} channels rely instead on $p_{T}^{b\bar{b}}>150$–160 GeV \cite{ATLAS:2018jvf}.
  
  \item CMS 8 TeV: $m_{jj}>250\;\text{GeV}$ (for events in Set A, which pass a dedicated signal trigger) or $>700\;\text{GeV}$ (for events in Set B, which pass a general-purpose VBF trigger) with $|\Delta\eta_{jj}|>2.5$–3.5 \cite{CMS:2015two}.
  
  \item CMS 13 TeV: \emph{Tight} paths use $m_{jj}>500$ (300) GeV for 2016 (2018) with $\Delta\eta_{jj}>4.1$ (3.5) and $\Delta\phi_{b\bar{b}}<1.6$ (1.9), where $phi$ is the azimuthal angle;
  \emph{Loose} paths keep $m_{jj}>250  (200)\;\text{GeV}$ for 2016 (2018) with $\Delta\eta_{jj}>2.3$ (1.5) and $\Delta\phi_{b\bar{b}}<2.1$ (2.8) \cite{CMS:2023tfj}.
\end{itemize}

Each analysis seeks 4 jets with 2 $b$-jets. Typical jet $p_{T}$ thresholds range from $50$ GeV in the ATLAS 8 TeV analysis up to
$110$ GeV in the CMS 13 TeV \emph{Tight} selection, while $b$‑jet candidates are subject to additional cuts. A detailed list of the 
Trigger and Offline selections for each analysis are presented on Tables~\ref{tab:2A8}-\ref{tab:2C13}.

All four \emph{VBF} $H\!\to\!b\bar b$ studies perform a simultaneous fit to the
reconstructed $m_{b\bar b}$ spectrum in several \emph{Signal Region (SR)} categories, but the way those categories are built differs:
ATLAS 8 TeV defines four BDT categories
  (plus a cut‑based cross‑check); the discriminant is trained on
  $\{m_{jj},\Delta\eta_{jj},p_{T}^{b\bar{b}}\}$ and a handful of soft‑QCD
  observables~\cite{ATLAS:2016mzy}.
ATLAS 13 TeV builds BDTs for the \emph{two‑central},
  \emph{four‑central} and $\gamma{+}$\emph{VBF} channels and fits the output in six
  bins, again driven mainly by $\{m_{jj},\Delta\eta_{jj},p_{T}^{b\bar{b}}\}$ \cite{ATLAS:2018jvf}.
CMS 8 TeV uses seven BDT regions, trained separately for
  the dedicated and general‑purpose \emph{VBF} triggers; the same three \emph{VBF}
  kinematics dominate the ranking~\cite{CMS:2015two}.
CMS 13 TeV applies a binary BDT in the \emph{Tight} data-sets
  and a three‑way \emph{(VBF/ggF/}$Z+$\emph{jets)} network in the \emph{Loose} sets,
  giving ten categories in total~\cite{CMS:2023tfj}.

\emph{Monte Carlo (MC)} samples are used for simulated events in analyses. For the signal \textsc{Powheg}+{\sc Pythia8} is used commonly in analyses \cite{ATLAS:2016mzy,ATLAS:2018jvf,CMS:2015two,CMS:2023tfj}:
ATLAS models $Z+$\emph{jets} with {\sc Sherpa}, while CMS generates $V+$\emph{jets}
and top backgrounds with {\sc MadGraph5}$\_$aMC$@$NLO+{\sc Pythia8};
\emph{Parton Distribution Function (PDF)} and scale variations are evaluated at \emph{Next-to-Leading-Order (NLO)}, and both 13 TeV analyses assess \emph{Parton Shower (PS)} systematics by swapping
{\sc Pythia8} for {\sc Herwig7}. Furthermore: ATLAS 8 TeV uses NNLO correctionswhereas the \emph{ggH} carry NNLO with an \emph{Next-to-Next-to-Leading Logarithmic (NNLL)} enhancement; ATLAS 13 TeV relies solely on NLO‑accurate \emph{Powheg} samples; CMS 13 TeV VBF achieves \emph{Next-to-Next-to-Leading-Order (NNLO)} QCD accuracy together with NLO electroweak corrections; and CMS 8 TeV establishes NNLO corrections together with NNLL resummation for gluon fusion. 

After the full category selections, the analyses retain
$\mathcal{O}(10^{2})$ \emph{VBF} $H\rightarrow b\bar{b}$ events in a \emph{luminosity} ($\mathcal{L}$) of 20.2 fb$^{-1}$ at ATLAS 8 TeV and in $19.8+18.3$ fb$^{-1}$ at CMS 8 TeV,
and about $\mathcal{O}(10^{3})$ in each of the Run‑2 data-sets analysed so far in upto 
$30.6$ fb$^{-1}$ at ATLAS 13 TeV and in $36.3+54.5$ fb$^{-1}$ at CMS 13 TeV \cite{ATLAS:2016mzy,CMS:2015two,ATLAS:2018jvf,CMS:2023tfj}. Consequently, the measurements reported here provide an important additional constraint on $\kappa_{V}$ and $\kappa_{b}$ when included in global fits that combine \emph{VH, ggF, ttH} and $H\rightarrow\gamma\gamma / ZZ / \tau\tau$ channels.


\subsection{Boosted Decision Tree strategies}
\label{sec:bdt_strategies}

To maximise separation of the electroweak $\mathrm{VBF}\,H\!\to\! b\bar b$ signal from backgrounds, all four \emph{VBF} $H\!\to b\bar b$ analyses employ BDT classifiers implemented in the {\sc Root} \emph{Toolkit for Multivariate Analyses (TMVA)}, but with experiment– and energy–specific training strategies \cite{ROOT:Release,Brun:1997pa}. 

In the CMS 8~TeV search, one BDT first assigns the two $b$- and two \emph{VBF}-tag jet candidates, improving signal efficiency by about 10$\%$, and subsequent BDT discriminants (separate for two data-sets) perform the signal–background separation. In the CMS 13~TeV measurement, separate binary BDTs (for \emph{Tight}) and a multiclass BDT (for \emph{Loose}) use \emph{VBF}-jet kinematics, $b$-tag scores, quark–gluon likelihoods, global momentum-balance, and additional jet-activity variables; a small fraction of Data ($\sim 3\%$) is included in training as a proxy for the dominant multijet background, with alternative splits showing $<2\%$ impact. 

The ATLAS 8~TeV analysis trains a single BDT against non-resonant background taken from $m_{b\bar{b}}$ sidebands, using jet-shape, additional-jet activity and \emph{VBF} topology variables while keeping inputs minimally correlated with $m_{b\bar{b}}$; its output defines four categories for the simultaneous $m_{b\bar{b}}$ fits. The ATLAS 13~TeV search trains channel-specific BDTs (two all-hadronic plus a photon-associated channel) whose outputs define multiple regions for a global $m_{b\bar{b}}$ fit separating non-resonant and $Z+$\emph{jets} backgrounds. Thus, while all analyses rely on BDT-based multivariate discrimination and BDT-defined categories, only CMS 8~TeV uses an additional jet-assignment BDT stage, CMS 13~TeV introduces a multiclass strategy. Both ATLAS analyses derive non-resonant background shapes by relying on Data sidebands (i.e., rather than solely using simulated multijet samples): For BDT training, the ATLAS 8~TeV analysis and the 13~TeV hadronic (\emph{two‑central} and \emph{four‑central}) channels use sideband Data as the background sample, whereas the 13~TeV \emph{photon} channel, lacking sufficient sideband statistics, employs a reweighted $\gamma+$\emph{jets} simulation sample.

The continuous BDT discriminant (most commonly named as $w$ in ATLAS and $D$ in CMS) outputs are partitioned into analysis categories whose numerical boundaries and yields are summarised in Tables~\ref{tab:3A8}-\ref{tab:3C13}:
\begin{itemize}
    \item ATLAS at 8~TeV defines four successive $w$ intervals ($-0.08<w\le 0.01$, $0.01<w\le 0.06$, $0.06<w\le 0.09$, and $w>0.09$), with total Data event-counts decreasing from $176{,}073$ (Category~$I$) to $6{,}493$ (Category~$IV$) while the expected \emph{VBF} yield falls more slowly from $39$ to $19$ events (See Table~\ref{tab:3A8}). 
    \item ATLAS at 13~TeV uses channel-specific $w$ selections: \emph{two‑central} ($w\ge -0.006$ vs.\ $w<-0.006$), \emph{four‑central} ($w>0.033$, $0.026<w\le 0.033$, $0.015<w\le 0.026$, $0.002<w\le 0.015$), and \emph{photon}$+$\emph{VBF} ($w>0.30$, $-0.05\le w\le 0.30$, $w<-0.05$); in each channel the higher-$w$ regions contain fewer Data events but a larger fraction of the quoted Higgs (\emph{VBF}+other) expectation, e.g.\ \emph{two‑central} region~I has $35{,}496$ Data with $340$ Higgs events vs.\ $95{,}802$ Data and $165$ Higgs events in region~II (See Table~\ref{tab:3A13}). 
    \item CMS at 8~TeV separates the two triggers on data-sets and defines non-overlapping $D$ intervals: Set~A ($-0.6<D\le 0.0$, $0.0<D\le 0.7$, $0.7<D\le 0.84$, $0.84<D\le 1.0$) and Set~B ($-0.1<D\le 0.4$, $0.4<D\le 0.8$, $0.8<D\le 1.0$), with Data counts shrinking from $546{,}121$ (A1) to $10{,}874$ (A4) and from $203{,}865$ (B5) to $15{,}151$ (B7) (See Table~\ref{tab:3C8}). Note that the discriminant is not explicitly named as $D$ in the CMS 8 TeV analysis paper \cite{CMS:2015two}, but the discriminant used is similar to $D$ in CMS 13 TeV \cite{CMS:2023tfj}.
    \item CMS at 13~TeV employs two structures: (i) \emph{Tight} categories using a single discriminant $D$ ($0.25<D<0.50$, $0.50<D<0.75$, and $D>0.75$) and (ii) \emph{Loose} categories defined by logical combinations of three multiclass outputs $(D_{\mathrm{VBF}}, D_{\mathrm{ggF}}, D_{Z})$; e.g.,\ $D_{\mathrm{ggF}}>0.55$ (G2), $D_{\mathrm{ggF}}<0.50$ and $D_{\mathrm{VBF}}>0.85$ (V2) in 2016; in both 2016 and 2018 Data the higher-$D$ (or higher-$D_{\mathrm{VBF}}$) bins have fewer total events and a larger fraction of the quoted \emph{VBF} expectation; e.g.,\ 2018 \emph{Tight} Category~3 ($D>0.75$) has $8{,}202$ Data events with $134.5$ \emph{VBF} vs.\ Category~1 ($0.25<D<0.50$) with $29{,}261$ Data and $89.5$ \emph{VBF} (See Table~\ref{tab:3C13}). 
\end{itemize}

Across all four analyses the tables show a consistent pattern: (a) monotonically shrinking data yields with increasing $w$ or $D$ thresholds, (b) comparatively slower reduction of expected \emph{VBF} yields, and (c) experiment-specific use of either a single scalar $w$ (in both ATLAS analyses and the CMS 8~TeV) or multiclass outputs to form mutually exclusive logical regions (in CMS 13~TeV).




\subsection{Background Processes and Suppression}

As a continuation of the multivariate methodology described in Sect.~\ref{sec:bdt_strategies}, both ATLAS and CMS use the BDT (or associated discriminant outputs) to suppress and constrain backgrounds. The continuous discriminant outputs are partitioned into ordered regions of increasing expected signal purity; the lowest-score (i.e., background-dominated) bins provide strong constraints on the dominant background normalisations and shapes that propagate to the higher-score (i.e., signal-enriched) bins. The significant backgrounds fall into three categories:

\paragraph{i) Non-resonant QCD multijet and jet--object misidentification.}
Pure QCD production of four or more jets constitutes the largest background. Signal-like candidates arise when (a) light-flavour or charm jets are mis-tagged as $b$-jets, or (b) genuine $b\bar b$ pairs originate from gluon splitting. Very small contributions can also come from electrons or photons that shower early and are reconstructed as jets. As the probability to pass the $b$-tagging algorithms by such electrons and photons is less than $0.01\%$, these are rounded under statistical uncertainties on the background contributions \cite{CMS-DP-2018-058,Hartman:2891112}. Because the QCD cross-section is many orders of magnitude larger than that of \emph{VBF} $H\!\to b\bar b$, both experiments continue to rely on data-driven methods to suppress such backgrounds \cite{CMS:2023vzh,ATLAS:2024pov}.

 In the ATLAS 8 and 13~TeV analyses, the non-resonant $m_{b\bar{b}}$ shape in each BDT (or channel) bin is taken directly from Data sidebands and fitted simultaneously with the signal components \cite{ATLAS:2016mzy,ATLAS:2018jvf}. The CMS 8~TeV analysis performs a simultaneous fit across BDT categories (Set A and Set B), using a common polynomial parameterisation of the multijet shape with per-category transfer factors to absorb residual differences \cite{CMS:2015two}. At 13~TeV, CMS trains binary (for \emph{Tight}) and multiclass (for \emph{Loose}) BDTs; a small prescaled fraction of post-selection Data (2.5\% for binary, 5\% for multiclass) is included solely to supply a realistic multijet sample for training/validation, while the final multijet shape is constrained by the sideband-rich, lower-output categories in the global $m_{b\bar{b}}$ fits \cite{CMS:2023tfj}. Additionally in the 13 TeV analyses central-jet vetoes and soft-activity requirements (track-based sums) further suppress colour-connected multijet and $ggF+$\emph{jets} contamination before the fit \cite{ATLAS:2018jvf,CMS:2023tfj}.

The lowest BDT intervals (Tables~\ref{tab:3A8}-\ref{tab:3C13}) are multijet-dominated and anchor the multijet normalisation in the global fit.

\paragraph{ii) Electroweak and other top.}
Processes containing genuine heavy flavour but distinct kinematics form the next tier can be suppressed by identifying these analyses through using MC samples and applying specific cuts. 
\begin{itemize}
    \item The largest electroweak heavy-flavour contribution in the signal window arises from $Z+$\emph{jets} (including $Z\to b\bar b$), with $W+$\emph{jets} typically subdominant once the $b\bar b$ mass requirement and tagging criteria are applied. \emph{Control Regions (CRs)} and $m_{b\bar b}$ sidebands normalise the $V+$\emph{jets} components; simulation systematics constrain extrapolations \cite{ATLAS:2024pov}. 
    \item Top-quark production ($t\bar t$ and single-top) contributes events where the $b$-jets from top decays plus additional radiation mimic the \emph{VBF} dijet topology; lepton-enriched control samples determine their normalisations. \item Diboson ($VV$) production and single-Higgs processes (other decay modes) are treated as minor backgrounds with cross-sections fixed to the SM expectations. 
\end{itemize}

In the most recent analyses, the shapes of these backgrounds are taken from \textsc{Powheg}+{\sc Pythia8} or \textsc{Sherpa} samples, with renormalisation/factorisation scales and PS tunes varied for systematics \cite{ATLAS:2024pov}. Residual $V$-tag mis-identification (boosted hadronic $V$ reconstructed as a large-$R$ jet without a true Higgs) and accidental small-$R$ dijet mass entries are included within the $V+$\emph{jets} and multijet estimates and constrained by the lower/intermediate BDT bins (Tables~\ref{tab:3A8}-\ref{tab:3C13}) \cite{ATLAS:2023jdk}.

\paragraph{iii) Consolidated minor or reducible backgrounds.}
Events where jets are mis-identified, in ways already constrained by the multijet or $V+$\emph{jets}, are effectively absorbed through the jet selection, $V$-tagging criteria, and the multijet transfer-factor methodology \cite{ATLAS:2023jdk}. Mis-tagged \emph{VH} contributions are suppressed through $V$-jet mass fits, whereas the progressive BDT binning (Tables~\ref{tab:3A8}-\ref{tab:3C13}) ensures any residual contributions, such as dijet resonances from QCD, are profiled within background-dominated intervals. \\

After the full selection and multivariate fit, QCD multijet processes remain the leading systematic uncertainty (up to $20\%$), followed by limited MC statistics in $t\bar t$ and $V+$\emph{jets} simulation \cite{ATLAS:2023jdk}. With these background definitions and data-driven constraints, the SR is dominated by signal-enriched events, enabling the reported \emph{VBF} $H\!\to b\bar b$ measurements to reach observed significances of $1.9-2.0\,\sigma$\cite{ATLAS:2018jvf,CMS:2023tfj}.

\section{Results}



\emph{VBF} production with subsequent $H\!\to b\bar b$ decay provides a clean electroweak topology but remains statistics- and background-limited. Present published measurements at $\sqrt{s}=8,13\,$TeV by ATLAS and CMS determine the \emph{VBF} signal strength $\mu_{\text{VBF}}$ with around $50\%$ precision for the higher energy analyses, leaving considerable room for improvement, compared to a few‑percent theory uncertainty on the SM prediction for the signal \emph{cross-section} ($\sigma$) and \emph{branching ratio} ($\mathcal{B}$): $\sigma_{\text{VBF}}\times\mathcal{B}(H\to b\bar b)$.

\subsection{Multivariate (BDT / classifier) response}
\label{subsec:bdt_response}

\begin{figure}[p]
    \centering
    
    \includegraphics[width=0.42\linewidth]{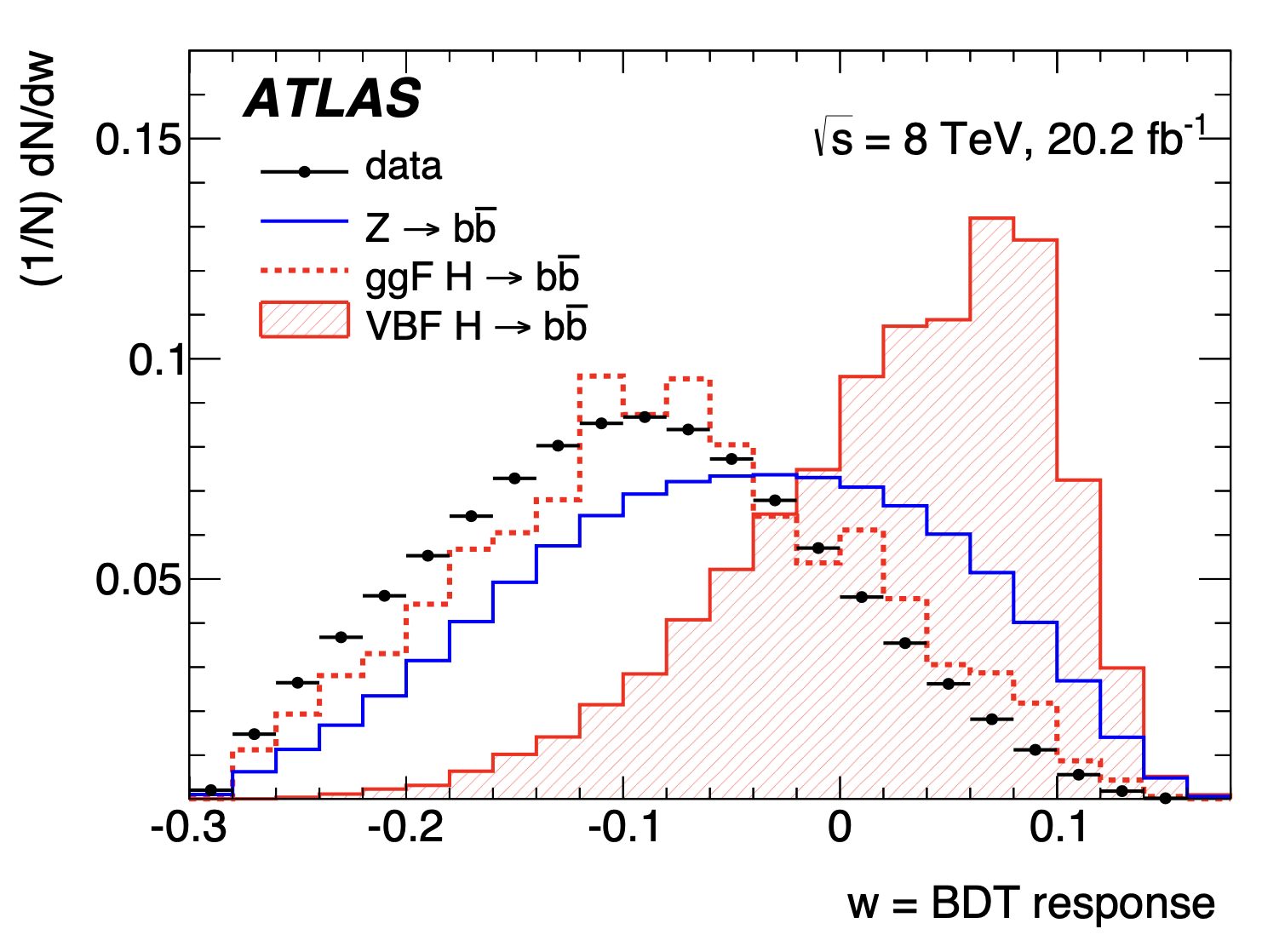}
    \includegraphics[width=0.41\linewidth]{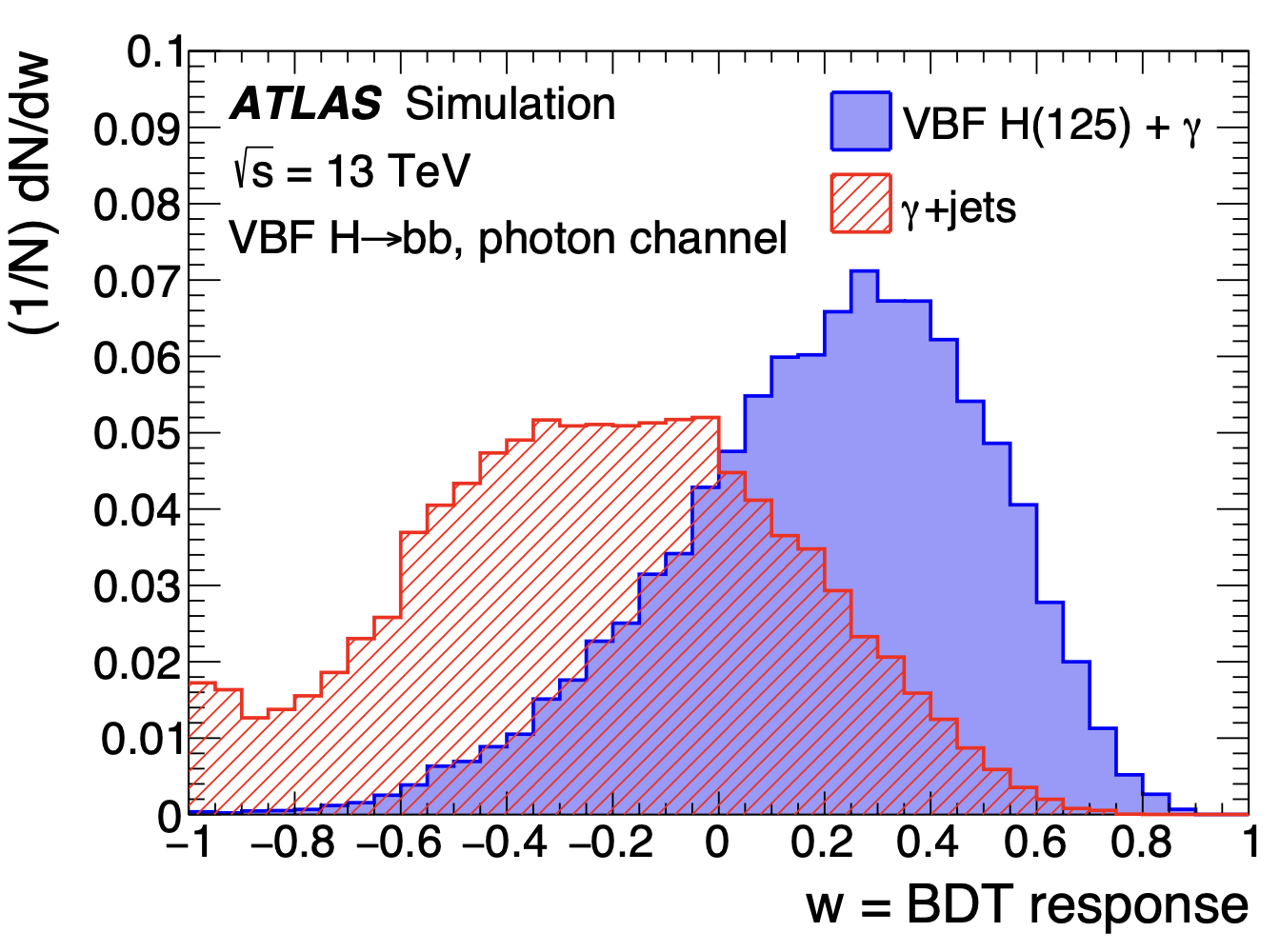}
    
    \includegraphics[width=0.84\linewidth]{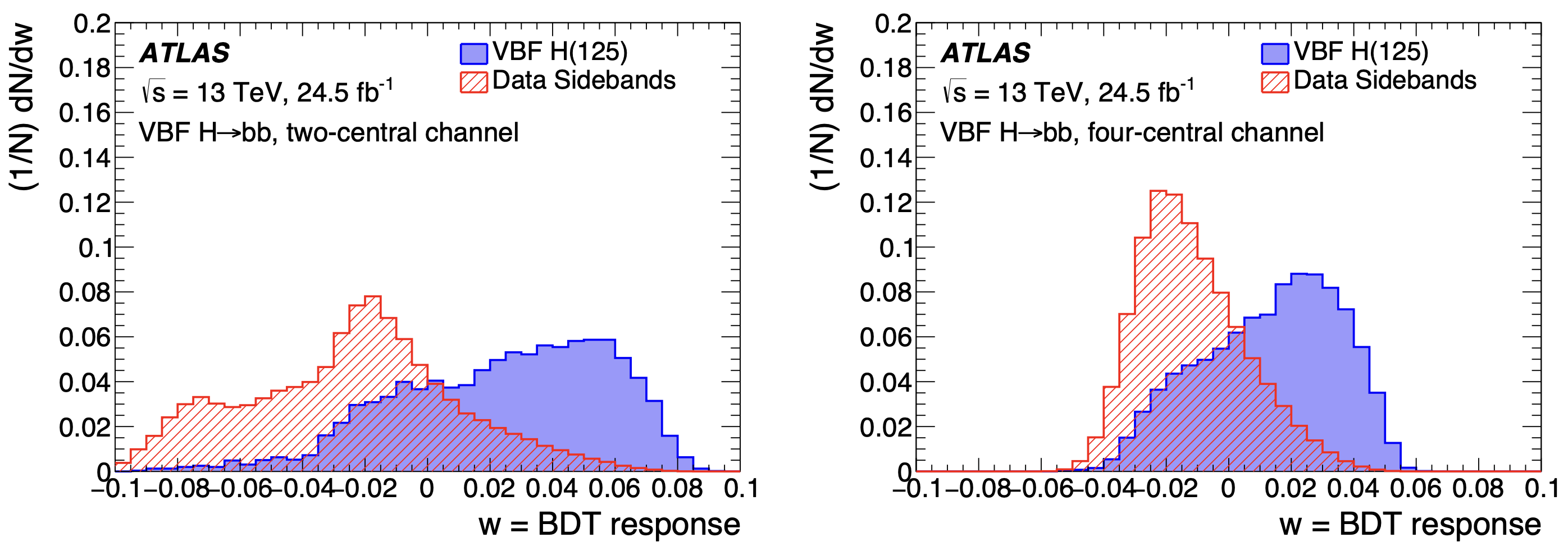}
    
    \includegraphics[width=0.94\linewidth]{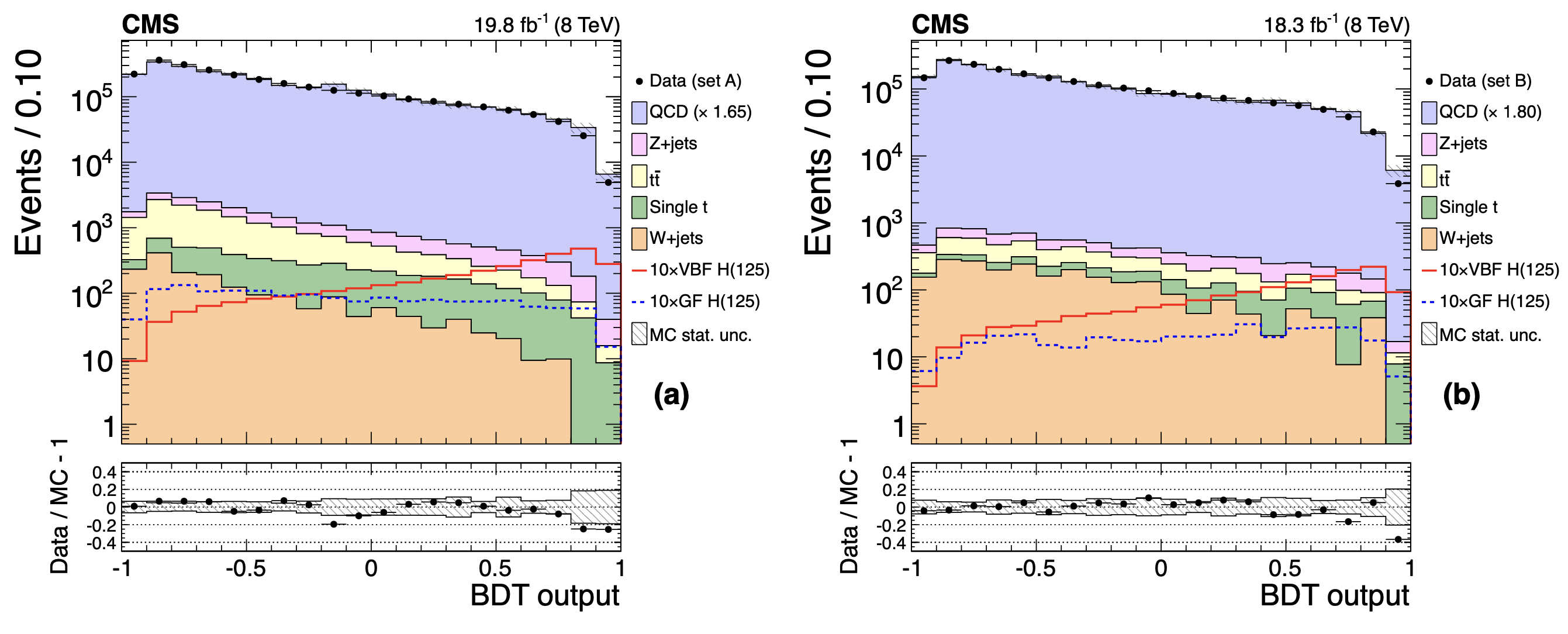}
    \includegraphics[width=0.8\linewidth]{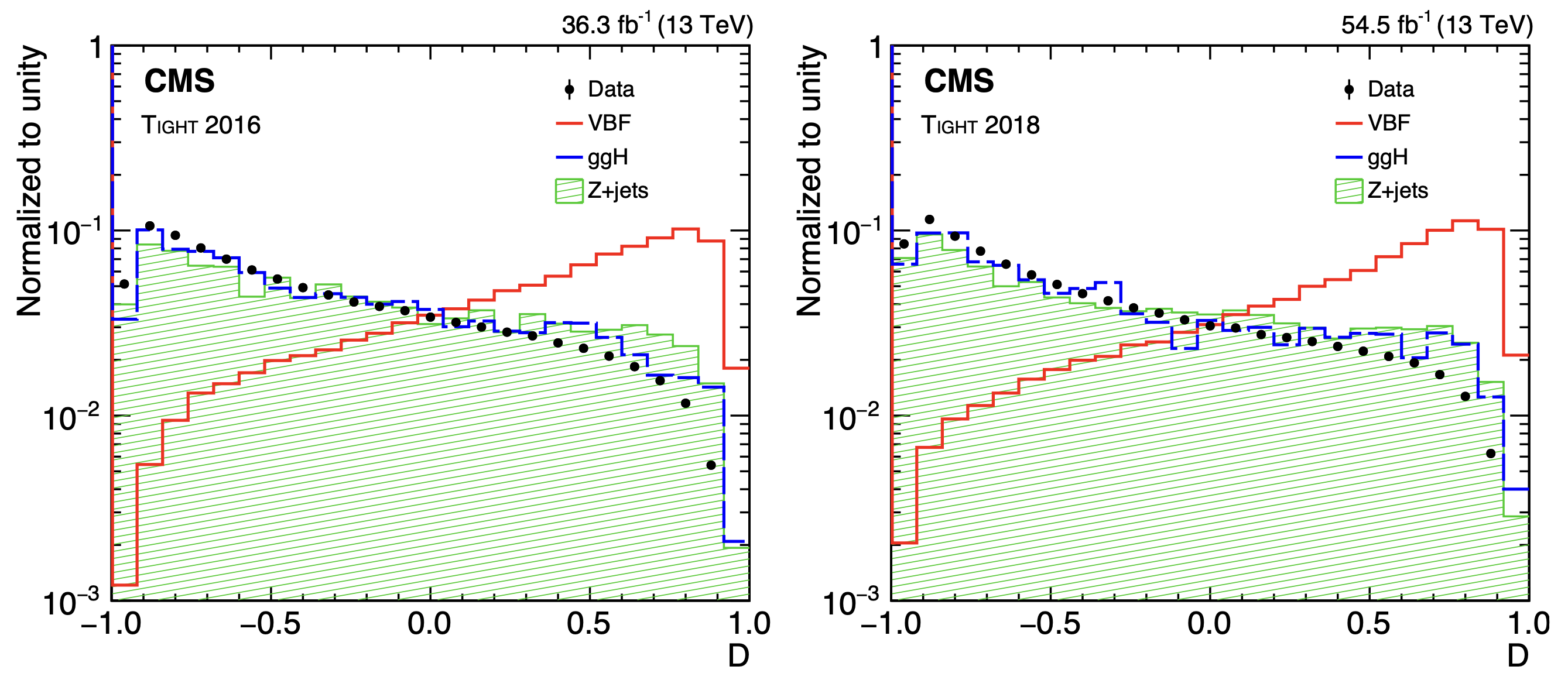}
     \caption{\scriptsize{
     \textbf{Multivariate discriminants that drive the \emph{VBF} \(\boldsymbol{H\!\to b\bar b}\) searches \cite{ATLAS:2016mzy,ATLAS:2018jvf,CMS:2015two,CMS:2023tfj}.} 
     In the unit–normalised histograms (all except CMS-8 TeV), the Higgs contribution is scaled up for visibility.
     These discriminants—from a simple quark–gluon likelihood to full BDT outputs—serve as the inputs or final selectors in each analysis.
     \textbf{Top-Left (1 panel)}:~ATLAS, 8 TeV: BDT response after pre‑selection. Data (points) are compared with stacked backgrounds; \emph{VBF}‑Higgs (red, filled), $ggF$‑Higgs (red, dashed) and \(Z\!\to b\bar b\) (blue) shapes are overlaid.  
     \textbf{Top-Right, Middle-Up (3 panels)}:~ATLAS, 13 TeV: BDT score \(w\) after the common \emph{VBF} pre-selection (blue, filled), displayed for the \emph{two‑central}, \emph{four‑central}, and photon\,+\,\emph{VBF} channels. The background histogram (red, filled) is taken from \(m_{b\bar b}\) side-bands (all-hadronic channels) or from the \(\gamma{+}\)jets simulation (\emph{photon} channel).
     \textbf{Middle-Down (2 panels)}:~CMS, 8 TeV: BDT output distributions for Set A and Set B. Data points are compared with stacked simulated backgrounds (LO QCD rescaled to match Data). The lower panel shows $(\text{Data}-\text{MC})/\text{MC}$ with the MC statistical band.
     \textbf{Bottom (2 panels)}:~CMS, 13 TeV: response of the dedicated \emph{VBF}-BDT in the \emph{Tight} category, shown separately for the 2016 (left) and 2018 (right) data-sets. Data points are dominated by QCD multijet events; coloured curves overlay the expected shapes for \emph{VBF} signal (red), $ggF$ spill-in (blue), and \(Z{+}\)jets (green, filled-in).
     }}
    \label{fig:two}
\end{figure}

After common kinematic preselection requiring two forward tagging jets and two central $b$ jets, each analysis partitions \emph{Signal (S)} and \emph{Background (B)} events using multivariate discriminants or classifier outputs whose shapes are shown in Fig.~\ref{fig:two}.

Overall, the multivariate categorisation:
\begin{enumerate}
  \item Establishes ordered purity slices increasing \emph{signal-to-background ratio} ($S/B$) from $\lesssim 10^{-3}$ at preselection to the percent level in the highest-score bins (illustrated by tightening categories in both experiments).
  \item Incorporates $Z+$\emph{jets} control categories (in CMS) to constrain the resonant component and reduce systematic pull on the signal measurement.
  \item Leverages auxiliary channels (in ATLAS \emph{photon}$+$\emph{VBF}) to exploit radiation patterns suppressing gluon-rich backgrounds.
\end{enumerate}

Across the four independent analyses, the collaborations follow the same strategic arc: 
construct a multivariate discriminant that tags the \emph{VBF} topology, 
slice that discriminant into purity‑ordered regions, and only then fit the $m_{b\bar{b}}$ spectrum for 
the Higgs signal.  In all cases, the Data events are overwhelmingly dominated by QCD multijet events, whereas the discriminant 
reveals a narrow peak where \emph{VBF} events concentrate, cleanly separated from the broad background band. 

\bigskip
\noindent
\textbf{ATLAS 8 TeV} (See Fig.~\ref{fig:two}, the Top-Left panel):\\
    The pre‑selected sample is dominated ($\gtrsim99\%$) by multijets,  but the BDT response rises steeply near $w\approx0.08$ for simulated \emph{VBF}–$H\!\to b\bar b$ and stays near zero for Data and for $Z$ and for gluon-fusion templates.  The spectrum is categorised by  minimising $\sqrt{S+B}/S$; only $6.5\times10^{3}$ of the $5.5\times10^{5}$ events end up in the highest‑purity bins in Category $IV$, yet the category carries the largest signal‑to‑background ratio and drives the combined fit.
    The separation between the broad, Data‑dominated background and the narrow signal‑like tail motivates the choice of a data‑driven fourth‑order Bernstein polynomial, fitted to the $m_{b\bar{b}}$ side‑bands, as the nominal description of the non‑resonant background; a fifth‑order form is kept only to estimate functional bias.  

\bigskip
\noindent
\textbf{ATLAS 13 TeV} (See Fig.~\ref{fig:two}, the Top-Right panel and the Middle-Up panels): \\
    A single BDT architecture is applied to three very different final states.  In all channels the  \emph{VBF}–H peak appears at positive scores and especially higher near $w\gtrsim0.04$, while side‑band Data cluster drops to around zero, allowing the analysis to define SR boundaries that both maximise sensitivity and retain sufficient $Z\!\to b\bar b$ events to pin down its normalisation to $<50\%$, thus enabling a stable \emph{VBF} extraction. Background shapes for the hadronic channels are taken directly from $m_{b\bar{b}}$ side‑bands, but the \emph{photon} channel must rely on a reweighted $\gamma+$\emph{jets} simulation, which is validated with iterative corrections, to stand in for the continuum, because too few Data events survive the isolation cuts.
    Once every object is identified as a jet (as in the \emph{two‑central} and \emph{four‑central}), the variables that distinguish quark‑initiated \emph{VBF} topology from multijet QCD vary only modestly. In contrast, the \emph{photon\,+\,VBF} panel spans the full $w\!\in[-1,1]$: the isolated‑photon trigger already suppresses much of the background, so the BDT can access a larger dynamical range. The lowest‑$w$ slice events are excluded from the SR set in the \emph{four‑central}, because their inclusion would dilute $S/\sqrt{B}$ and blow up the floated $Z\!\to b\bar b$ normalisation beyond the analysis cap of $1.5\times$ the SM prediction, whereas the \emph{two‑central} keeps the same slice as a single control‑like region that stabilises the side‑band template. 

\bigskip
\noindent
\textbf{CMS 8 TeV} (See Fig.~\ref{fig:two}, the Middle-Down panels):\\
    CMS trains two BDTs with identical inputs but a different trigger selection (Set A and Set B). For each, the leading‑order QCD prediction is rescaled until the stacked  background matches the Data; even after that, the \emph{VBF} signal, drawn $\times10$ for visibility, forms a peak near $D\simeq1$ that is absent in the other backgrounds. After the single scale factor is applied, the ratio panels $(\text{Data}-\text{MC})/\text{MC}$ stay within $\pm20\%$ across the full BDT range and always lie inside the MC‑statistical band, confirming the shape choice for the QCD template. Since the MC‑statistical band already dominates the uncertainty, the figure also signals that further progress at 8 TeV would have required larger simulation samples rather than more sophisticated discriminants, which has seemingly informed the 13 TeV upgrade.
    A diagnostic purity of $S/B$ anchors each of the seven BDT categories: in the best bin (Set~A, Cat. 4) $S/B \simeq 1.7\%$, while lower‑purity slices are an order of magnitude smaller. These ratios are reported only for validation; the final result still comes from one global likelihood fit, without any $S/(S{+}B)$ weighting. 

\bigskip
\noindent
\textbf{CMS 13 TeV} (Fig.~\ref{fig:two}, see the Bottom panels):\\
    In the \emph{Tight} samples a binary BDT is trained on \emph{VBF} vs.\ all other backgrounds by using only 5\,\% of the Data for training.  The residual 95\,\% that is taken as a proxy for QCD remains flat,  whereas simulated \emph{VBF} events peak around $D\approx0.8$.  The same pattern recurs in 2016 and 2018 Data despite different triggers, validating the choice of three \emph{Tight} and fifteen \emph{Loose} categories used later in the Higgs fit, while also clearing any year based bias.
    Optimising the boundaries for $S/\sqrt{B}$ in each slice increases the expected QCD sensitivity, and the categorisation of $Z$ and $ggF$ supply \emph{in situ} CRs that anchor the normalisations of those backgrounds to within $\sim30$\,\% (e.g., in $Z\rightarrow b\bar{b}$ 13 TeV analyses), thereby preventing correlations with the \emph{VBF} yield from inflating the overall uncertainty.    

\bigskip
\noindent
Taken together, these figures make two points.  First, real LHC Data in the 
$b\bar b+jj$ final state (with two $b$-tagged jets traced back to the Higgs decay and additional two jets of any type within the selected VBF topology) are indeed dominated by QCD, as shown by the broad, featureless plateau of 
the discriminant in every experiment.  Second, and crucial for Higgs physics, the \emph{VBF} signal inhabits 
a markedly different region of multivariate phase space: quark‑like tag jets, large rapidity gaps, 
high $m_{jj}$, and suppressed soft activity.  That separation produces a pronounced peak in the BDT 
distributions, allowing to isolate a \emph{VBF}‑enriched subset with minimal QCD contamination.  
By studying the Higgs boson this way we can efficiently probe electroweak couplings, jet‑flavour 
observables and colour‑flow effects with a precision unattainable in the gluon‑fusion channel.  
Thus, \emph{VBF} is not merely an alternative production mode but it can also support Higgs physics 
free from the ambiguity of QCD‑driven backgrounds.

All four plots consequently justify the use of data-driven methods either by anchoring MC templates to inclusive yields, by sculpting CR transfer factors directly from side‑bands, or by reserving the vast majority of events as an \emph{in situ} template. This Data‑first strategy sustains the analyses against theory‑scale uncertainties and leaves the \emph{VBF} signal free to emerge in a region of multivariate phase space, where the hadronic approach is the cleanest.

\subsection{Signal strengths and uncertainties}
\label{subsec:signal_results}

\begin{figure}[p]
    \centering
    \includegraphics[width=.86\linewidth]{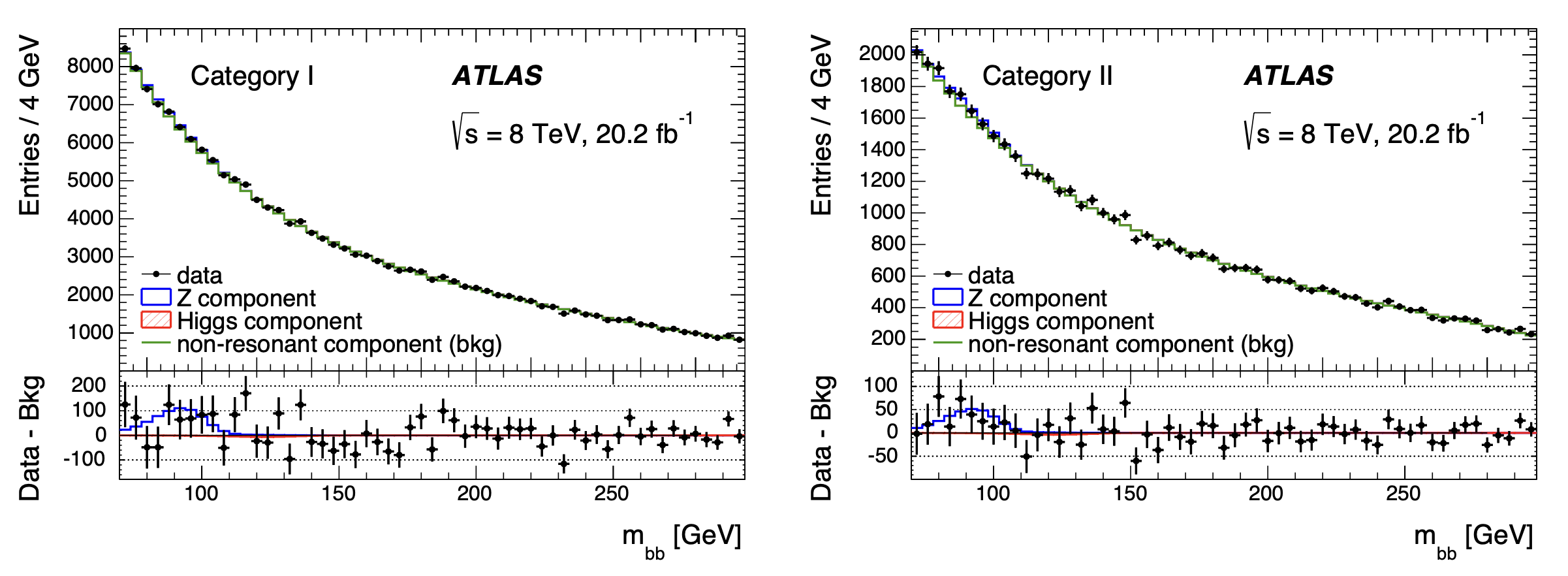}
    \includegraphics[width=.86\linewidth]{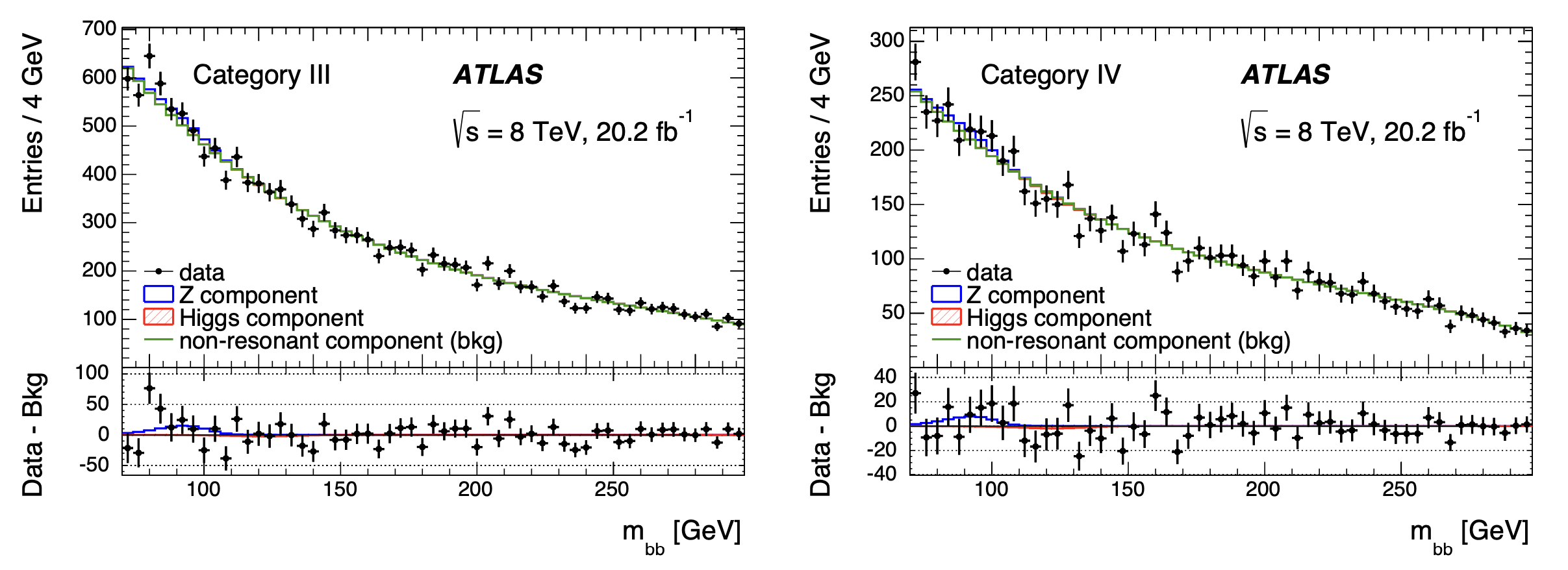}
    \includegraphics[width=0.78\linewidth]{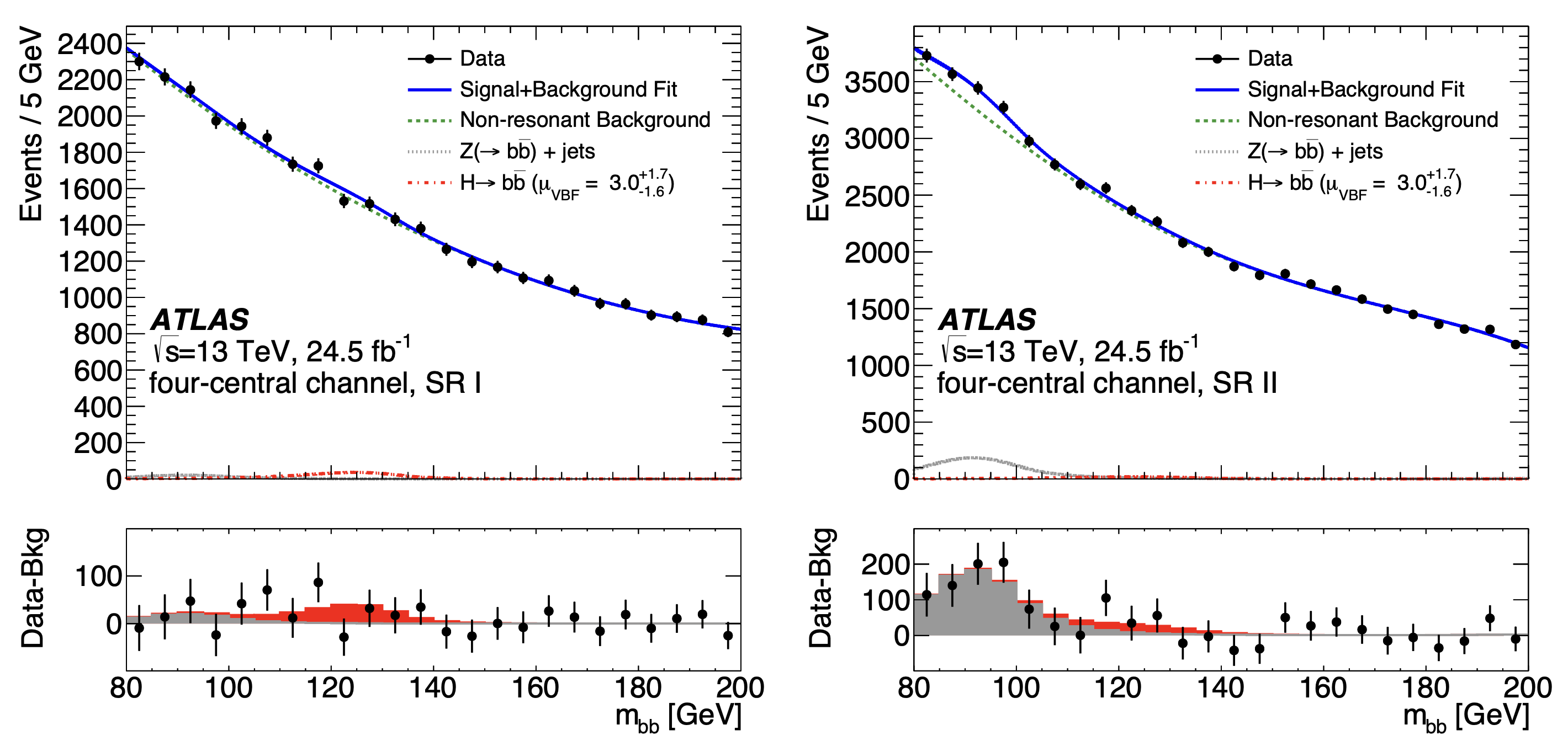}   

    \includegraphics[width=0.78\linewidth]{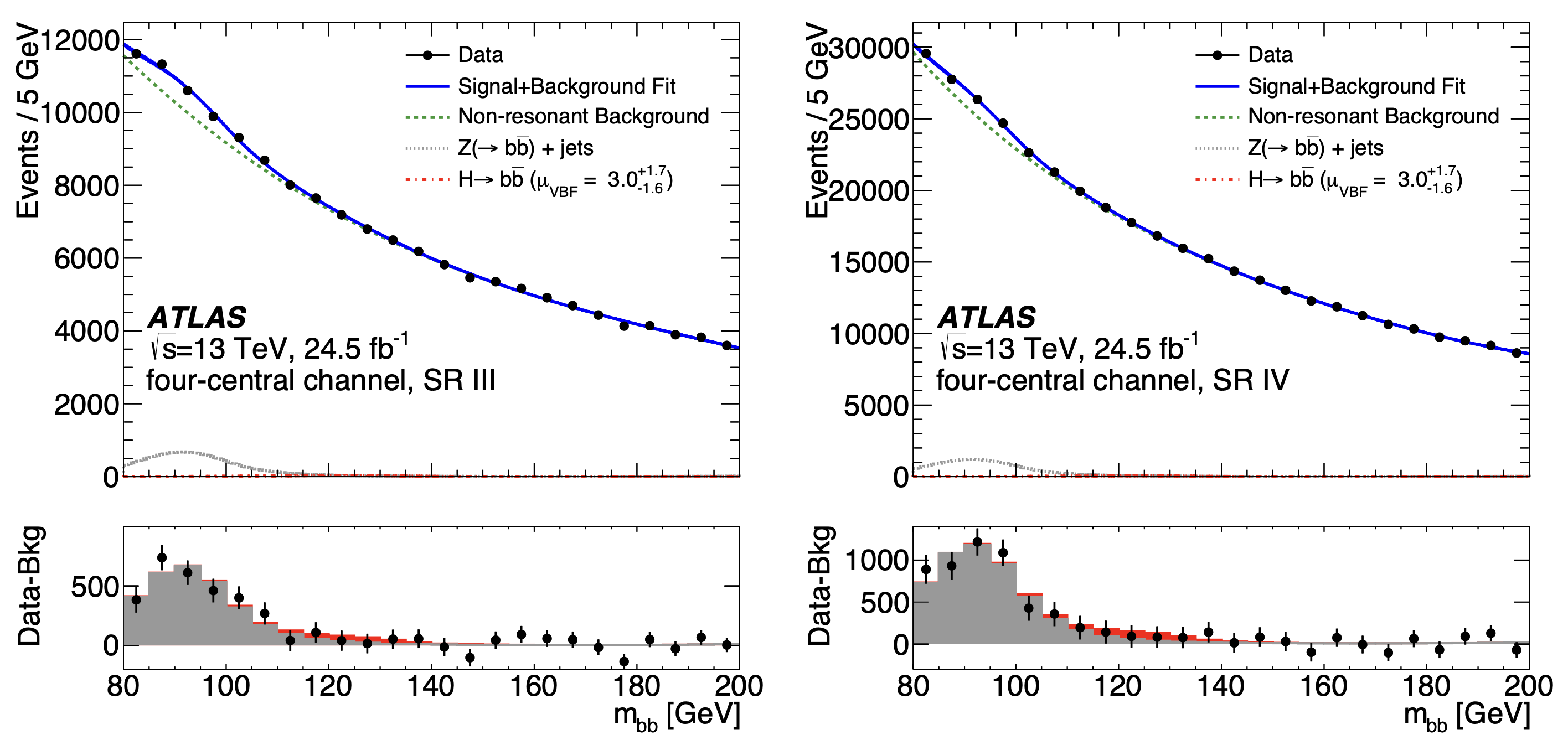}

     \caption{\scriptsize{
     \textbf{Four-tier multivariate \(m_{bb}\) distributions in ATLAS \emph{VBF} $H\!\to b\bar b$ analyses \cite{ATLAS:2016mzy,ATLAS:2018jvf}.}  Invariant–mass ($m_{bb}$) spectra of the two leading $b$‑jet candidates are shown in the most sensitive event classes of the ATLAS \emph{VBF} $H\!\to\!b\bar b$ searches.  Only statistical uncertainties are shown.
     \textbf{Top row (4 panels)}: ATLAS 8 TeV results of the profile‑likelihood fit to the $m_{bb}$ distributions in the four multivariate BDT categories \textbf{I–IV}.  Data points are compared with the fitted non‑resonant continuum (filled histogram), the resonant $Z\!\to b\bar b$ background (grey dotted), and the combined $VBF+ggF$ Higgs‐boson signal hypothesis (normalised to the fitted strength).  The lower insets show the Data after subtraction of the non‑resonant background together with the fitted $Z$ and Higgs contributions.
     \textbf{Bottom row (4 panels)}: ATLAS 13 TeV $m_{bb}$ distributions in the four \textit{four‑central} \emph{Signal Regions} \textbf{SR I–IV} after the global profile‑likelihood fit. The total post-fit (blue) is shown with the continuum background (green), the $Z\!\to b\bar b$ (grey, dotted), and  the fitted \emph{VBF} Higgs signal (red, dash-dotted). The residual panels display the Data minus the continuum background together with the fitted resonant components.
     }}
    \label{fig:thr-one}
\end{figure}

\begin{figure}[p]
    \centering
    \includegraphics[width=.8\linewidth]{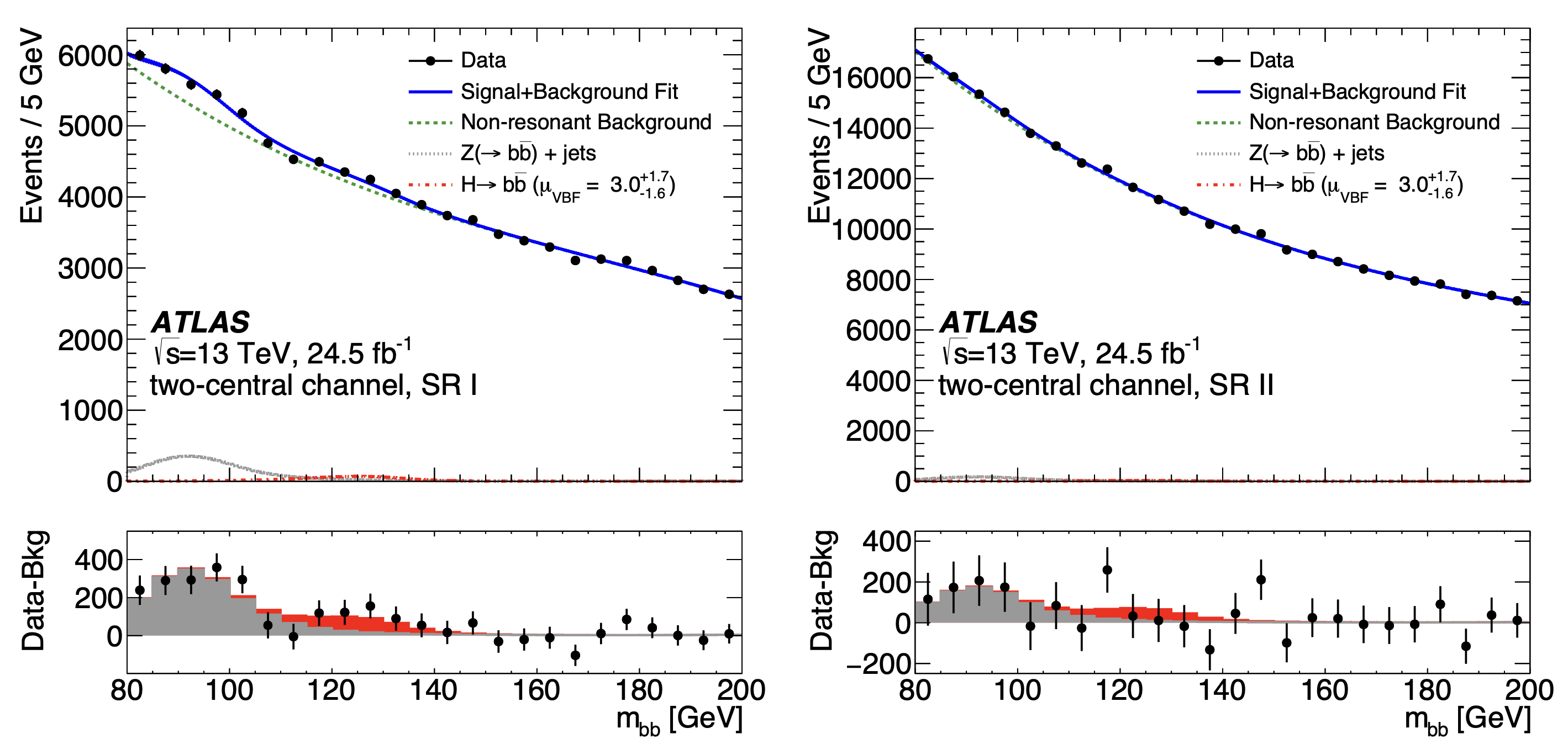}
    \includegraphics[width=.8\linewidth]{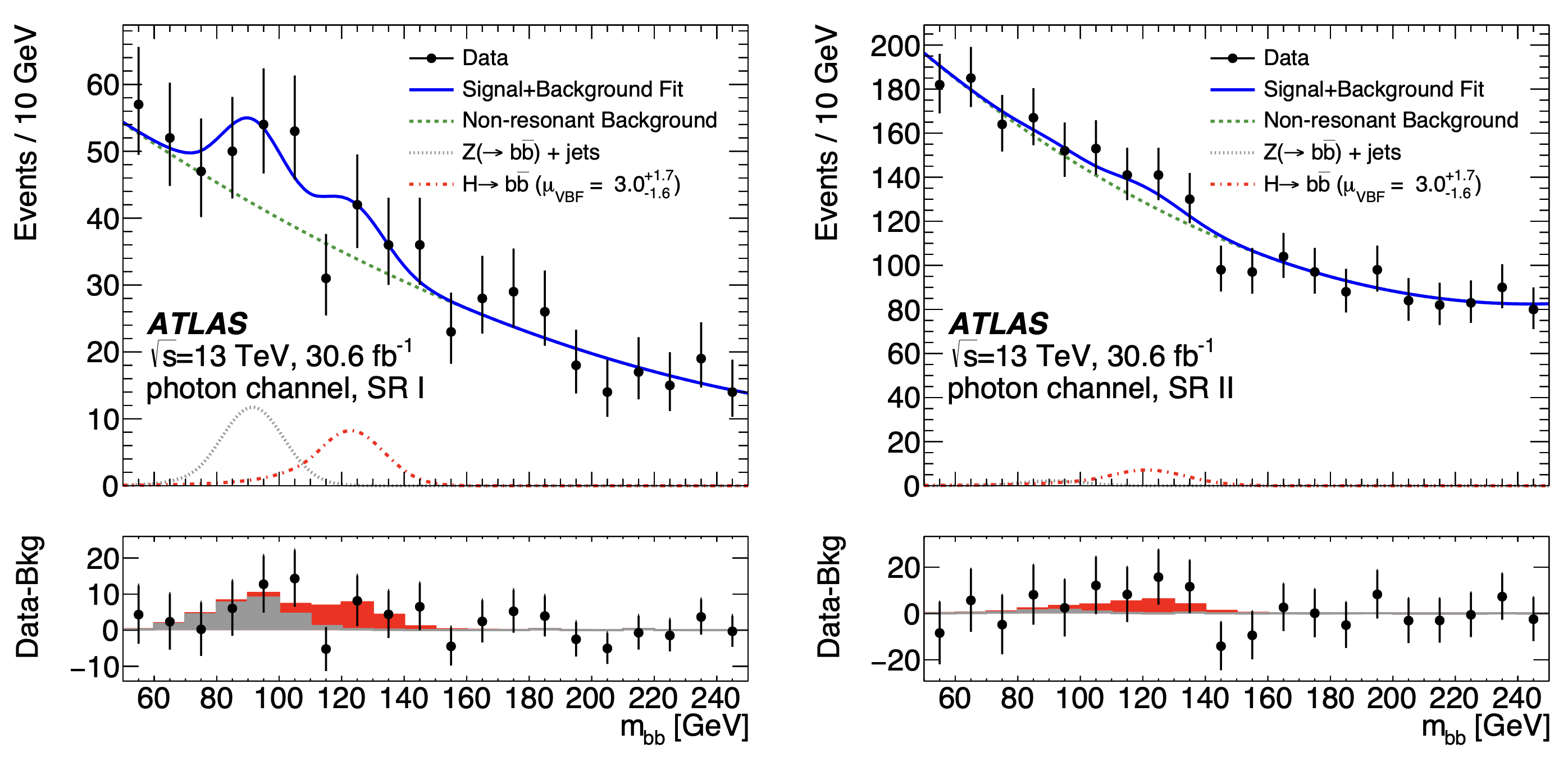}
    \includegraphics[width=0.4\linewidth]{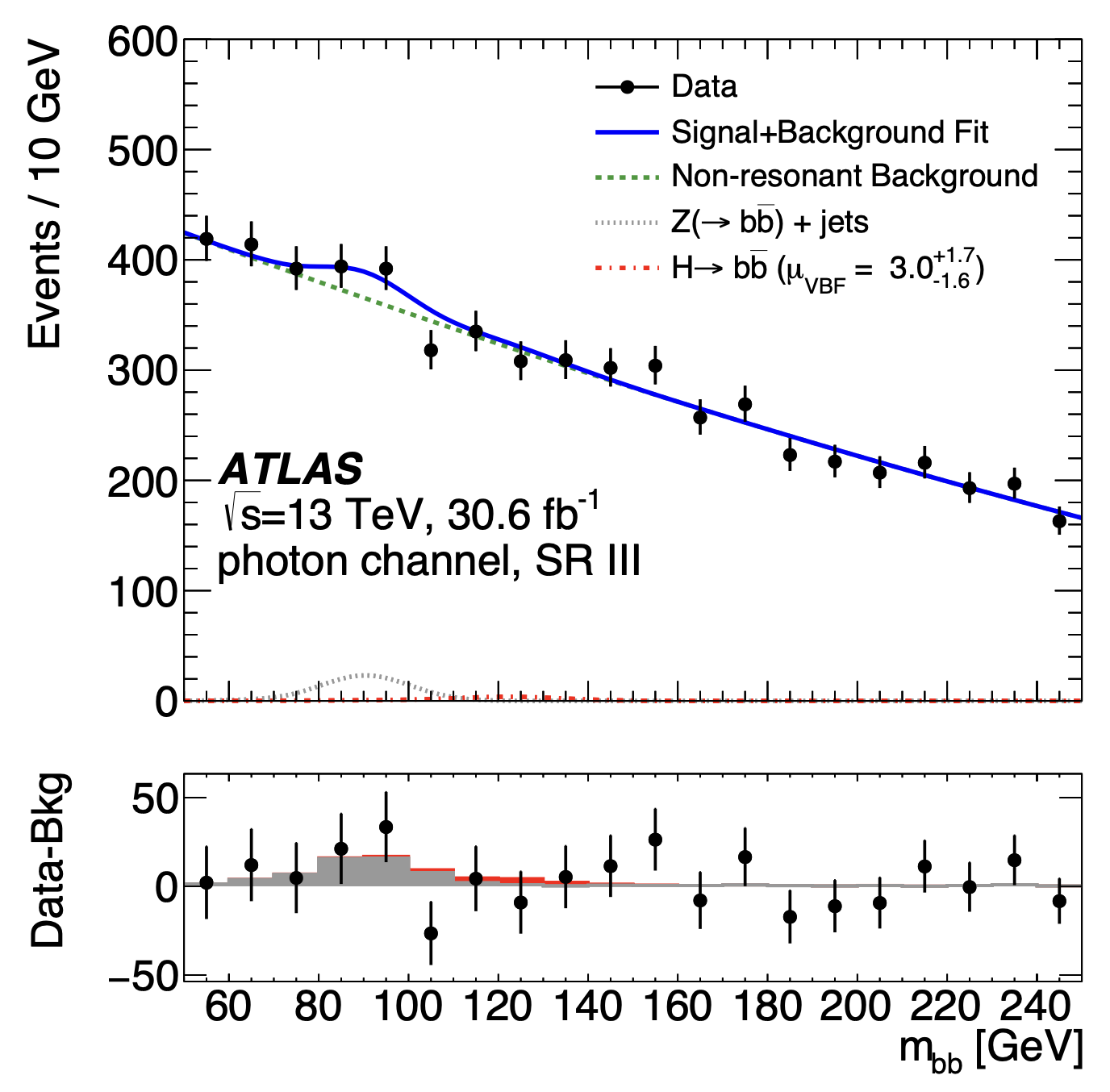}\includegraphics[width=.44\linewidth]{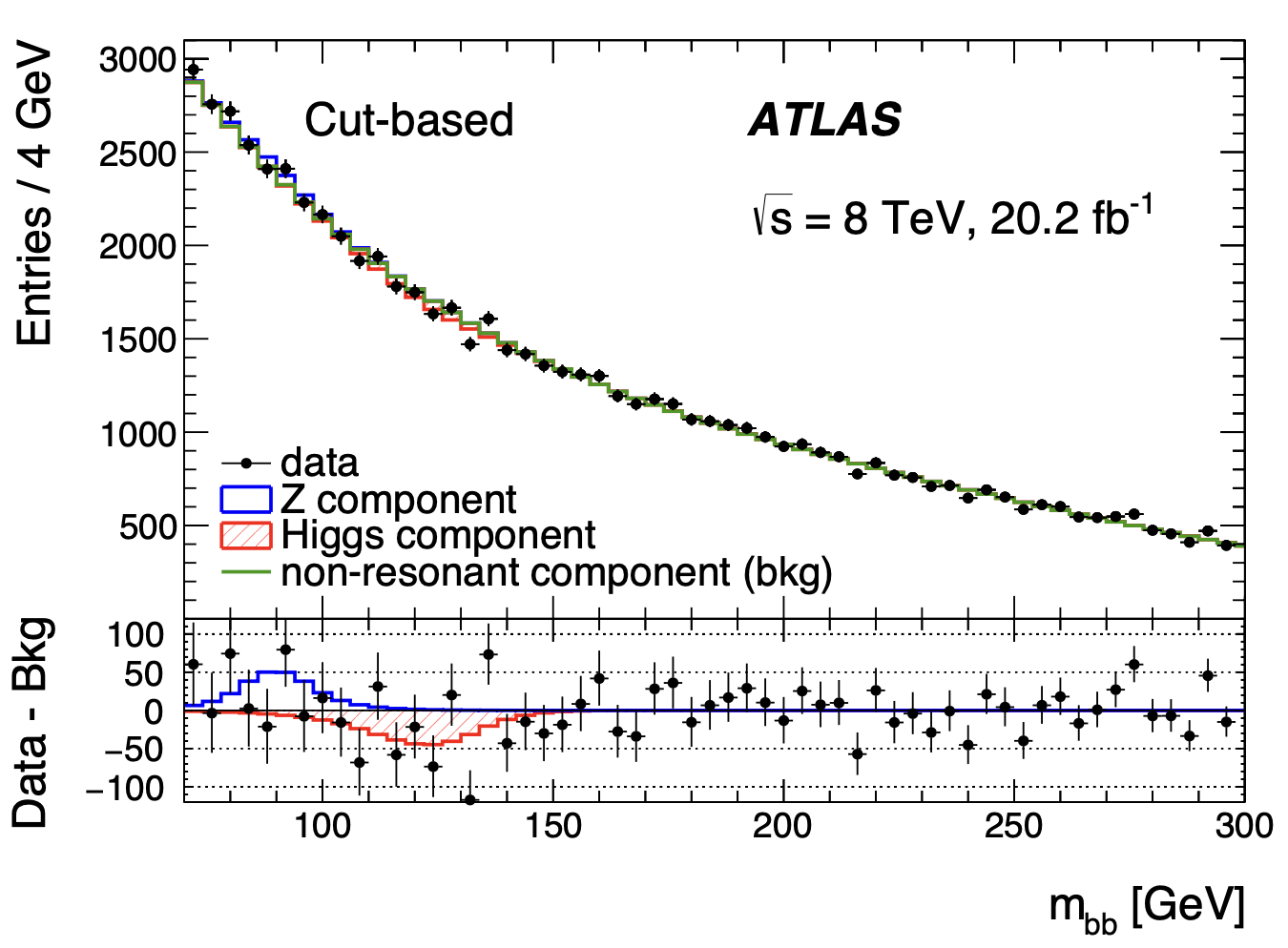}

     \caption{\scriptsize{
     \textbf{Other \(m_{bb}\) distributions in ATLAS \emph{VBF} $H\!\to b\bar b$ analyses \cite{ATLAS:2016mzy,ATLAS:2018jvf}.} Invariant–mass ($m_{bb}$) spectra of the two leading $b$‑jet candidates are shown in the most sensitive event classes of the ATLAS \emph{VBF} $H\!\to\!b\bar b$ searches. Only statistical uncertainties are shown. The solid blue curve shows the total post‑fit model; the dashed green curve the continuum background; the grey dotted curve the $Z\!\to b\bar b$ component; and the dash‑dotted red curve the fitted \emph{VBF} Higgs signal. The residual panels display the Data minus the continuum background together with the fitted resonant components.
     \textbf{Top row (2 panels)}: 
     ATLAS 13 TeV 24.5 fb$^{-1}$ $m_{bb}$ distributions in the four \textit{two‑central} \emph{Signal Regions} \textbf{SR I–II} after the global profile‑likelihood fit.  
     \textbf{Middle row and bottom-left (3 panels)}: ATLAS 13 TeV 30.6 fb$^{-1}$ $m_{bb}$ distributions in the four \textit{photon} \emph{Signal Regions} \textbf{SR I–III} after the global profile‑likelihood fit. 
    \textbf{Bottom-right (1 panel)}: ATLAS 8 TeV 20.2 fb$^{-1}$ $m_{bb}$ distribution with events selected by the cut-based analysis. 
     }}
    \label{fig:thr-two}
\end{figure}

\begin{figure}[p]
    \centering
    \includegraphics[width=0.84\linewidth]{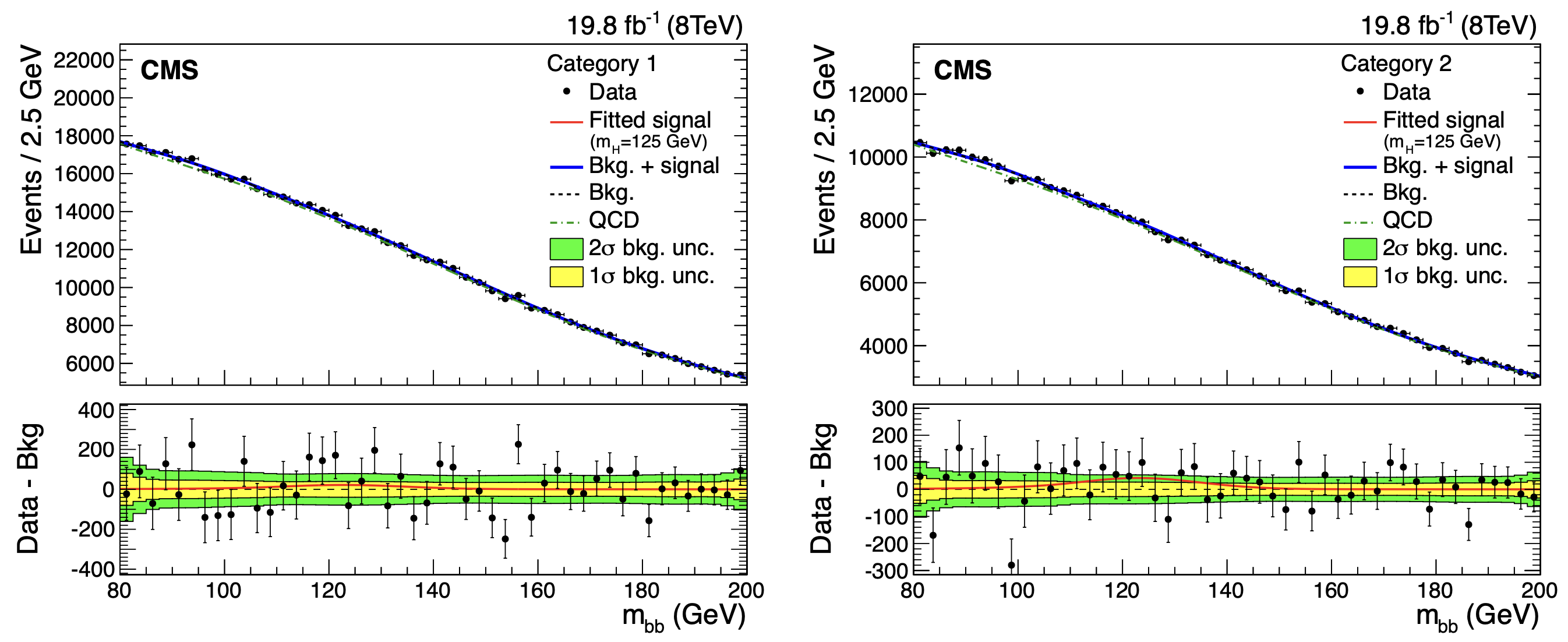}
    \includegraphics[width=0.84\linewidth]{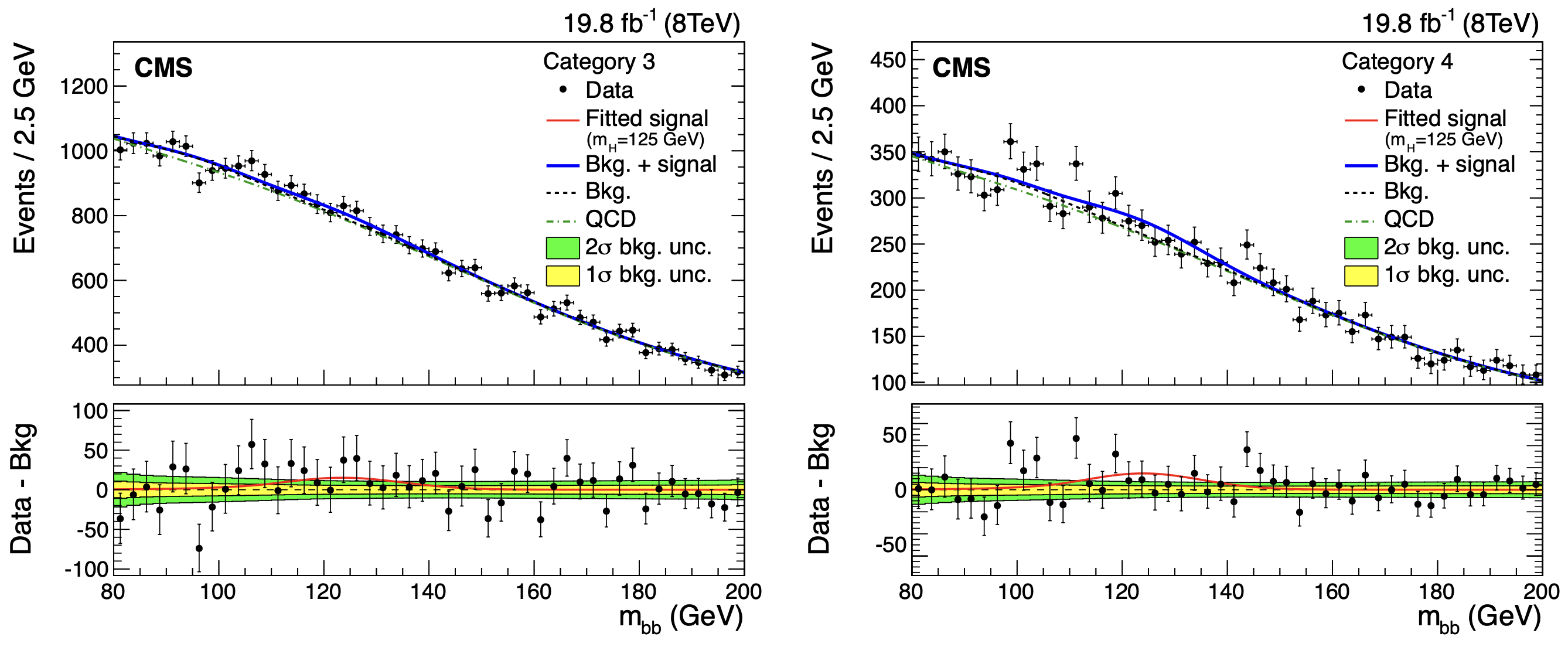}
    \includegraphics[width=0.84\linewidth]{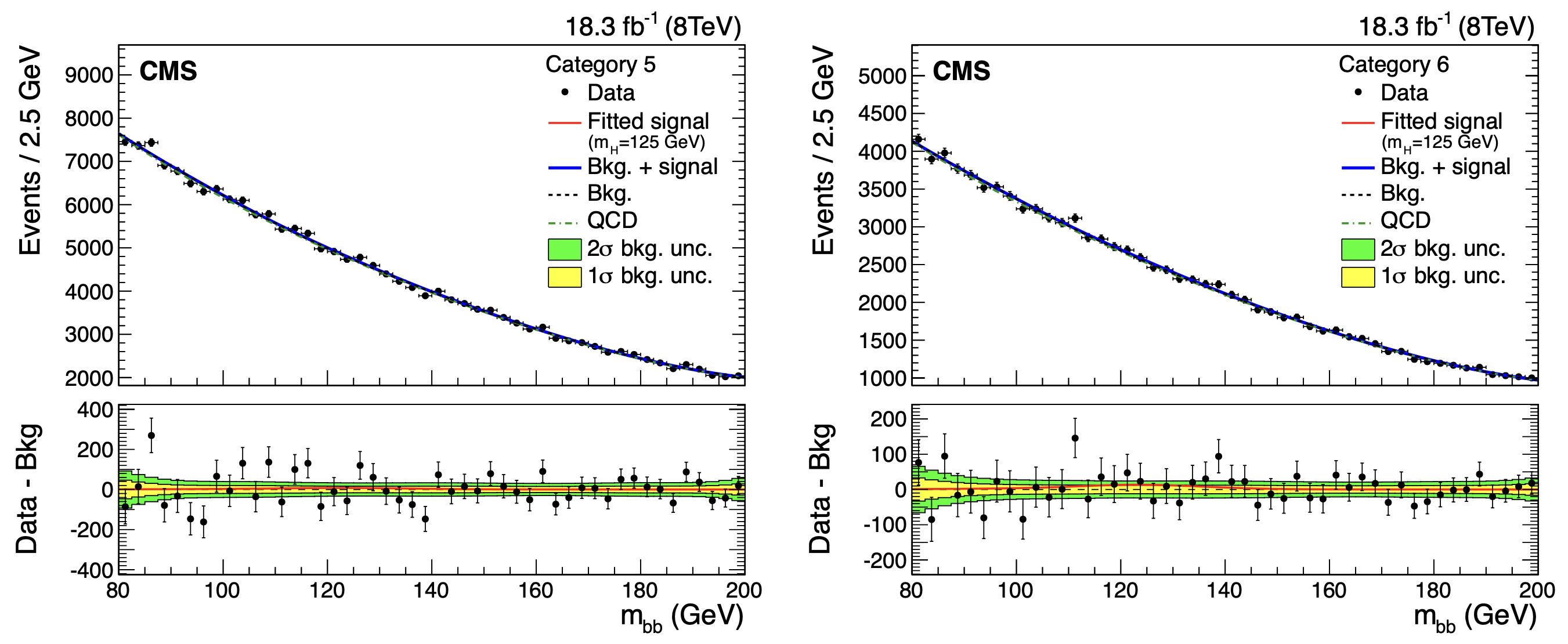}
    \includegraphics[width=0.42\linewidth]{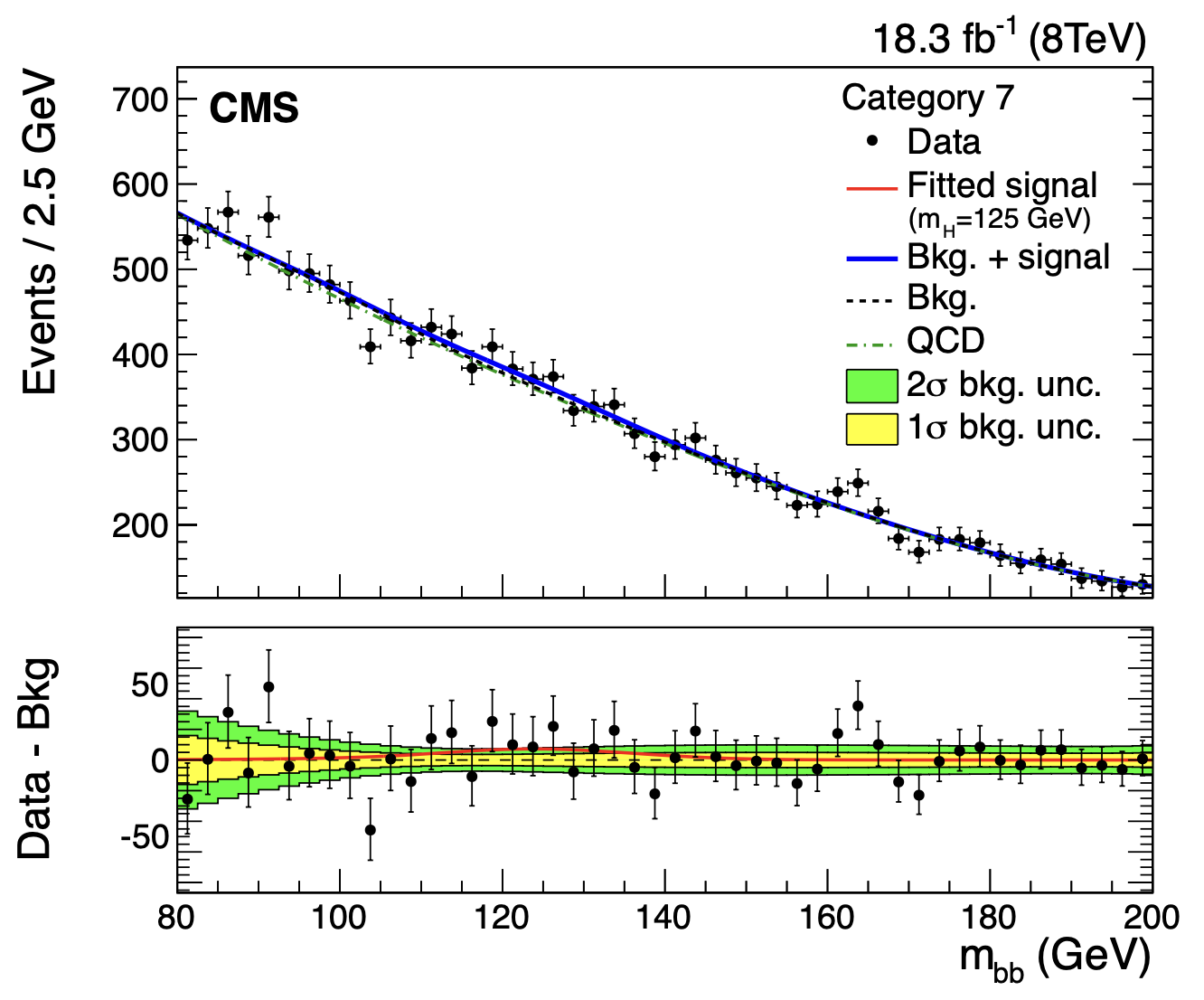}
     \caption{\scriptsize{
     \textbf{\(m_{bb}\) distribution in the seven most sensitive CMS\,8 TeV BDT categories in \emph{VBF} $H\!\to b\bar b$ analysis \cite{CMS:2015two}.} The invariant mass of the two leading \(b\)-jet candidates is shown for each category of the CMS \emph{VBF} \(H\!\to b\bar b\) search.  Statistical uncertainties only. 
     In every panel, black points represent Data; the solid blue curve shows the post‑fit signal + background model (including \emph{VBF} and $ggF$ \(H\!\to b\bar b\)); the dashed dark green curve is the fitted background‑only prediction; the dash‑dotted light green curve indicates the isolated QCD component.  Lower insets display the Data (black) and the fitted signal (red) after subtraction of the fitted background together with the \(1\sigma\) (yellow band) and \(2\sigma\) (green band) post‑fit uncertainty envelopes.
     \textbf{Top row (4 panels)}: Categories \textbf{1–4} obtained with the dedicated \emph{VBF} trigger (Set A, \(19.8\;\text{fb}^{-1}\)).  \textbf{Bottom row (3 panels)}: Categories \textbf{5–7} selected with the general‑purpose \emph{VBF} trigger (Set B, \(18.3\;\text{fb}^{-1}\)).
     }}
    \label{fig:thr-thr}
\end{figure}

\begin{figure}[p]
    \centering

    \includegraphics[width=1\linewidth]{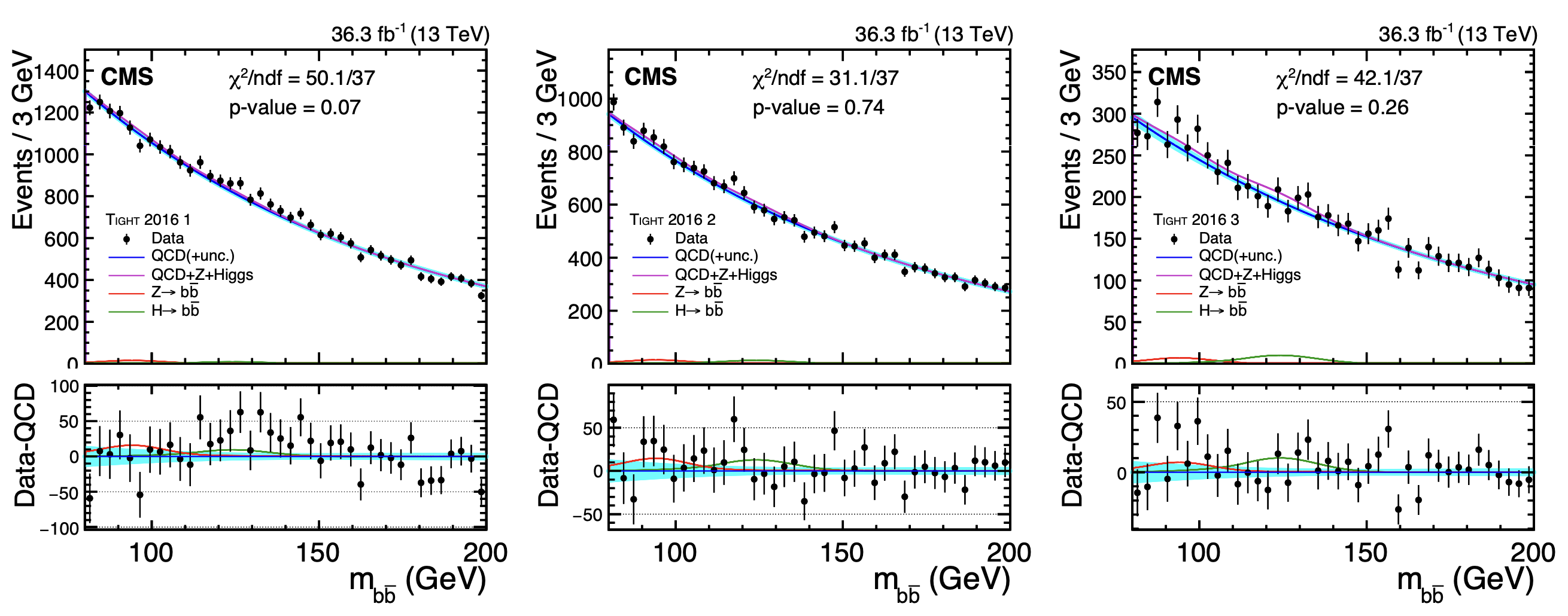}

    \includegraphics[width=1\linewidth]{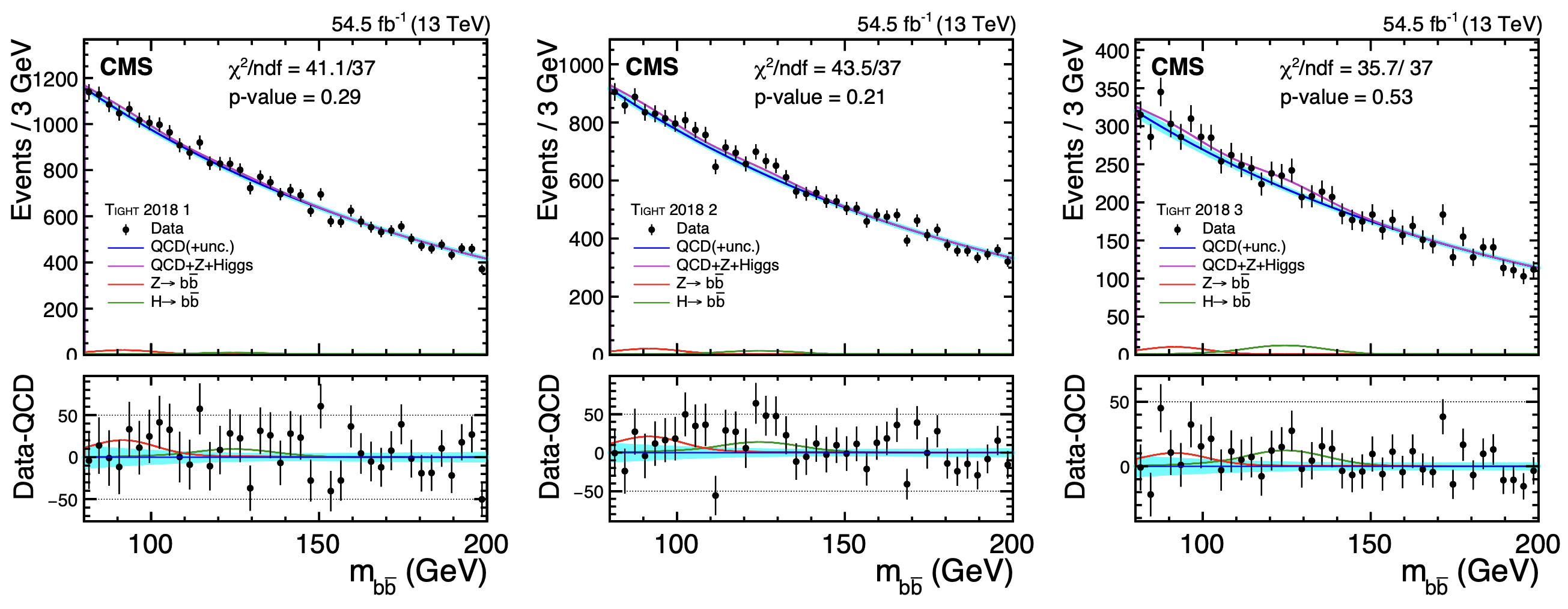}

    \includegraphics[width=0.66\linewidth]{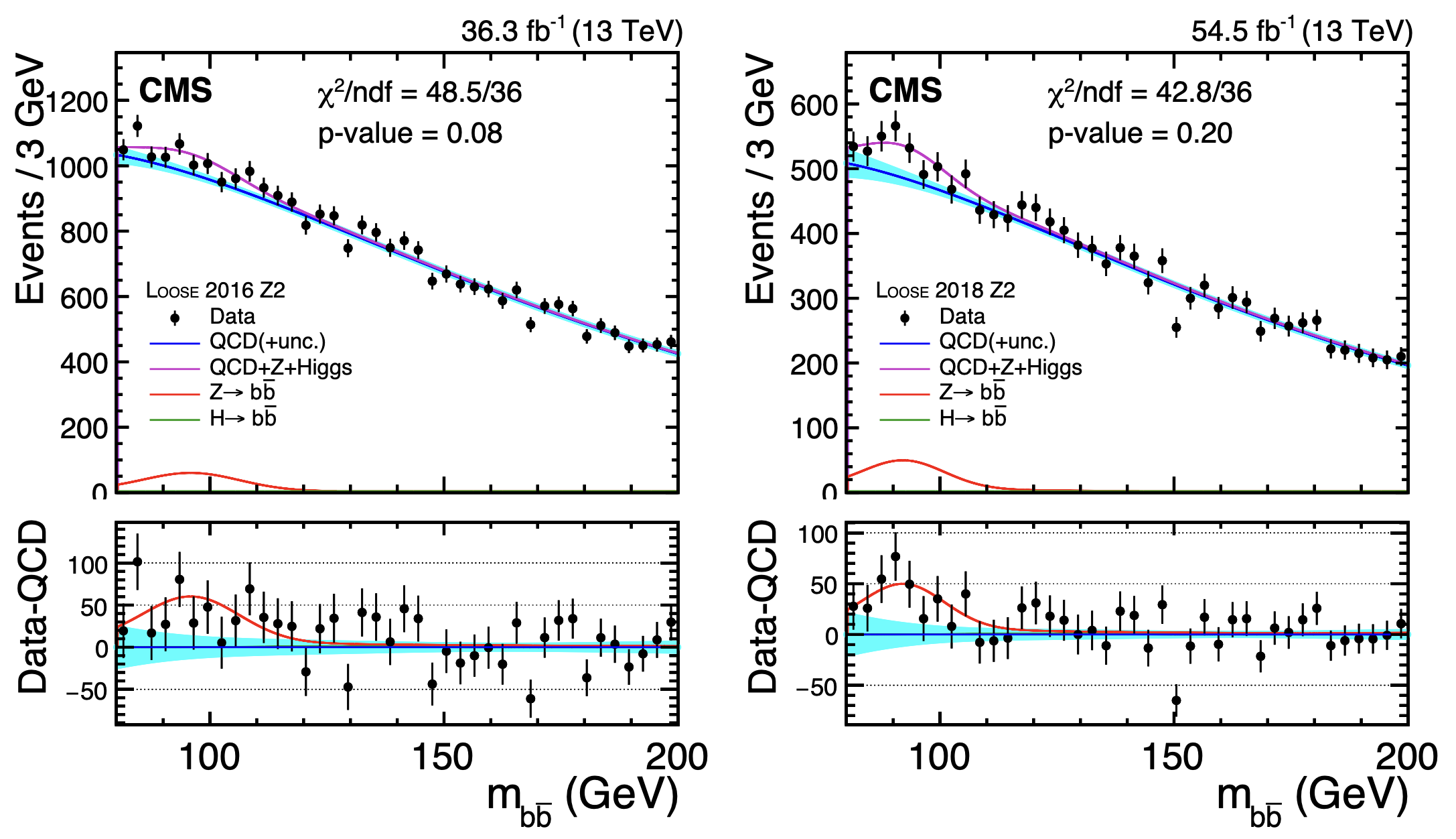}
    \includegraphics[width=0.33\linewidth]{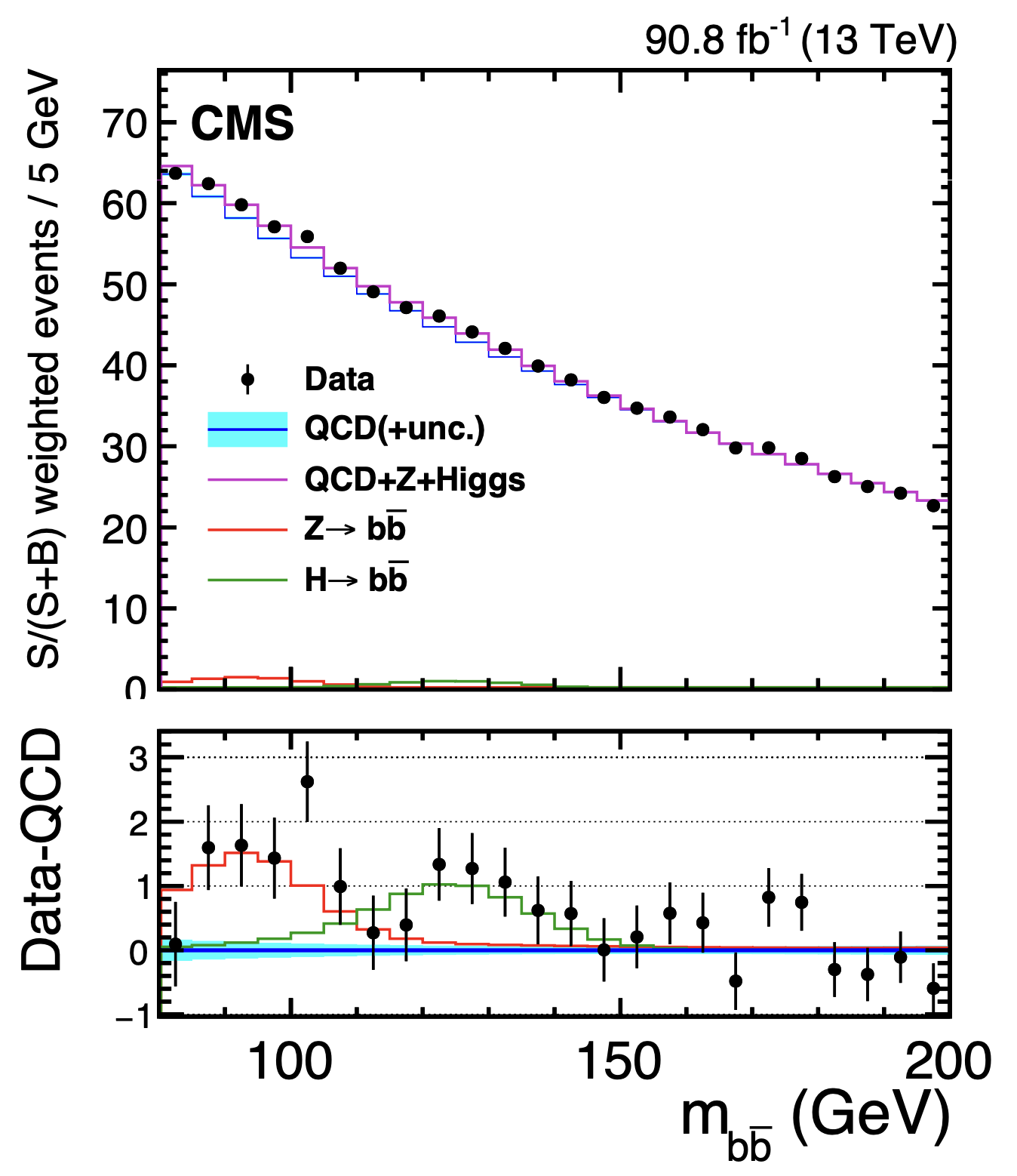}
    
     \caption{\scriptsize{
     \textbf{$m_{bb}$ distributions in the CMS 13 TeV \emph{VBF} $H\!\to b\bar b$ analysis \cite{CMS:2023tfj}.} 
     In all panels black points are Data; the blue solid curve is the fitted non-resonant background (dominated by QCD multijets) with its $\pm1\sigma$ uncertainty shown as a cyan band; the magenta curve is the total signal\,+\,background model (QCD\,+\,$Z\!\to b\bar b$\,+\,$H\!\to b\bar b$). After subtracting the non-resonant component, the lower \emph{Data-QCD} plots display the resonant pieces: red for $Z\!\to b\bar b$ and green for $H\!\to b\bar b$. 
     \textbf{Top (3 panels)}: \emph{Tight}~2016 categories \textbf{1–3} (left–right), with a luminosity of $36.3\,fb^{-1}$.  
     \textbf{Middle (3 panels)}: \emph{Tight}~2018 categories \textbf{1–3}, shown with the same convention, with a luminosity of $54.5\,fb^{-1}$.  
     \textbf{Bottom-right (2 panels)}: \emph{Loose}~Z2 categories for 2016 and 2018 (left to right), with a luminosity of $36.3\,fb^{-1}$ and $54.5\,fb^{-1}$. These categories are used to constrain the $Z\!\to b\bar b$ background. 
     For the first eight panels, each pad lists the $\chi^{2}/\text{ndf}$ and corresponding $p$‑value to quantify the goodness of fit. 
     \textbf{Bottom-left (1 panel)}: $m_{bb}$ spectrum obtained by combining all 18 categories with an $S/(S{+}B)$ weight (where $S$ is the total $H\!\to b\bar b$ yield from \emph{VBF} and $ggF$ (i.e., $ggH$)). 
     }}
    \label{fig:thr-fou}
\end{figure}

\noindent
In all four studies the signal yield is obtained exclusively from a \emph{simultaneous, binned maximum‑likelihood fit} to the di‑$b$‑jet invariant mass spectra, $m_{b\bar b}$, across multiple event categories that are pre‑selected with multivariate discriminants to enhance the \emph{VBF}–to–background ratio.  The fit floats only the $m_{b\bar b}$ templates for the Higgs resonance, the irreducible $Z\!\to b\bar b$ peak, and the smoothly falling multijet continuum, while systematic effects (jet‑energy calibration, $b$‑tagging, theory, etc.) are incorporated as profiled nuisance parameters constrained by side‑bands and CRs.  The result is expressed through the signal‑strength modifier
\[
\mu \;=\; \frac{\sigma_{\text{obs}}}{\sigma_{\mathrm{SM}}},
\]
which equals unity for the Standard Model prediction; values of $\mu$ extracted from the fits therefore quantify any excess or deficit of Higgs boson events seen in the $m_{b\bar b}$ distributions.  When gluon‑fusion production is floated independently (as in the CMS 13 TeV analysis) its yield is treated as an additional resonance in the same $m_{b\bar b}$ fit, anti-correlated with the \emph{VBF} component, but the signal is still extracted directly from the invariant mass fits of the two $b$-tagged jets. Figures~\ref{fig:thr-one}-\ref{fig:thr-fou} present the di‑$b$-jet invariant mass spectra, $m_{b\bar{b}}$, in the most sensitive event classes of the \emph{VBF}‑enriched $H\!\to\!b\bar b$ searches.\\

The \emph{VBF} signal strengths obtained by the four analyses are (See Table~\ref{tab:6} for the corresponding observed and expected \emph{Upper Limits (UL)} and significances):
\[
\mu_{\text{VBF}} =
\begin{cases}
-0.8 \pm 2.3 & \text{(ATLAS 8 TeV)}\\
3.0^{+1.7}_{-1.6} & \text{(ATLAS 13 TeV)}\\
2.8^{+1.6}_{-1.4} & \text{(CMS 8 TeV)}\\
1.01^{+0.55}_{-0.46} & \text{(CMS 13 TeV)}
\end{cases}
\]
.



\bigskip
\noindent
\textbf{ATLAS 8 TeV} (See Fig.~\ref{fig:thr-one} and Fig.~\ref{fig:thr-two}):\\
The final $m_{b\bar b}$ distributions are presented in BDT‑ordered SRs for the \emph{multivariate analysis (MVA)} in Fig.~\ref{fig:thr-one}: top 4 panels and for the \emph{cut-based analysis} in Fig.~\ref{fig:thr-two} bottom-right panel.  
\begin{enumerate}
  \item  Using $\mathcal{L}=20.2\;\mathrm{fb}^{-1}$ of Data, the profile‑likelihood fit returns a combined signal strength of $\mu = -0.8 \pm 2.3$ for MVA (and $\mu = -5.2 \pm \sim 4.6$ for cut-based). Consequently, a 95 \% \emph{Confidence Level (CL)} upper limit of $4.4$ is set, in line with expectation.
  \item After subtracting the fitted continuum (See the $Z$ component (blue) in \emph{Data-Bkg} plots on each panel), the residual spectra reveal a clean $Z\!\to b\bar b$ peak (\mbox{$m_{b\bar b}\simeq91$ GeV}) and no distortions elsewhere, demonstrating that the polynomial description of multijet QCD and the transfer of the shape between side‑ and signal‑bands work within statistical limits. 
  \item Also visible on the distributions is the the Higgs component (red) drop on all plots (most visibly on the cut-based analysis) around the mass of the Higgs (\mbox{$m_{b\bar b}\simeq125$ GeV}). This is an artifact of the post-fit normalisation as the best-fit signal strength is negative (and has a higher negative value in the cut-based case).
  This is merely a statistical fluctuation and the profiling of nuisance parameters rather than any physical suppression of the Higgs line shape.
  \item  Moving from the least (Category $I$) to the most (Category $IV$) sensitive region on the MVA distribution (Fig.~\ref{fig:thr-one}: top 4 panels), the $S/B$ ratio grows by almost an order of magnitude, validating the BDT topology variables that enhance forward‑jet kinematics and central $b$‑tag multiplicity.  
  \item Key uncertainty contributors are the \emph{Jet Energy Scale (JES)} at $15\%$ and the \emph{Jet Energy Resolution (JER)} at $4\%$; trigger plus $b$‑tag calibration, $10$–$20\%$; $Z+$\emph{jets} scale factors, $25\%$ correlated (up to $40$–$50\%$ uncorrelated); and the non‑resonant background parametrisation, where $\pm1.0\%$ is from function choice and $\pm1.7\%$ from sideband statistics. 
\end{enumerate}

\bigskip
\noindent
\textbf{ATLAS 13 TeV} (See Fig.~\ref{fig:thr-one} and Fig.~\ref{fig:thr-two}):\\
The Run~2 analysis splits the Data into \emph{four‑central} channel as shown in Fig.~\ref{fig:thr-one} vs. \emph{two‑central} and \emph{photon} channels as shown in Fig.~\ref{fig:thr-two}; the spectra shown correspond to the most sensitive SRs of each channel.
\begin{enumerate}
  \item Using $\mathcal{L}=24.5\;\mathrm{fb}^{-1}$ of Data for the \emph{four‑central} and \emph{two‑central} channels and $\mathcal{L}=30.6\;\mathrm{fb}^{-1}$ of Data for the \emph{photon} channel, the simultaneous fit yields $\mu_{\mathrm{VBF}} = 3.0^{+1.7}_{-1.6}$ and $\mu_{\mathrm{incl.}} = 2.5^{+1.4}_{-1.3}$, driven mainly by the \emph{four‑central} regions.
  \item Although the signal measurement carries an observed significance of $1.9\,\sigma$, it remains compatible with both the Standard Model expectation and statistical fluctuations.  
  \item The $Z$ mass peak (gray) for confirmation and the Higgs peak (red; the peak is positive as the signal value is positive) are present in the \emph{Data-Bkg} plots, as expected.
  \item The photon‑associated category (in Fig.~\ref{fig:thr-two} middle two panels an the bottom-left panel) carries very little $ggF$ contamination on the \emph{VBF} signal and displays a shape consistent with background (SR III is almost unaffected by the artificial Higgs peak as well), thus tightening the constraint on the \emph{VBF} cross section when included in the global fit.
  \item Roughly $60\%$ of the error budget is purely statistical ($\pm0.9$), while the leading systematic sources are the non‑resonant background shape, $\pm1.2$ on~$\mu$; $Z+$\emph{jets} normalisation, $\pm0.5$; JES/JER (suppressed through calibration); and the Higgs modelling near $\pm0.3$.
\end{enumerate}

\bigskip
\noindent
\textbf{CMS 8 TeV} (See Fig.~\ref{fig:thr-thr}):\\
As CMS adopts seven BDT categories (four in Set A, three in Set B), the $m_{b\bar b}$ distributions illustrate the most signal‑rich regions.
\begin{enumerate}
  \item Using  $\mathcal{L}=19.8+18.3\;\mathrm{fb}^{-1}$ (Set A + Set B) of Data, the combined fit reports $\mu = 2.8^{+1.6}_{-1.4}$ with an observed significance of $2.2\,\sigma$, a mild upward fluctuation relative to the $0.8\,\sigma$ expectation.
  \item Even though the $Z\rightarrow b\bar{b}$ isn't plotted separately, the 91 GeV $Z$ mass bump is visible in Set A's signal-plus-background: $S+B$ (blue; Fig.~\ref{fig:thr-thr}: top 4 panels). The fitted signal (red) remains smooth against the falling QCD template (plus the Higgs mass peak, also distinguishable on $S+B$), whose universality across categories was verified with a linear transfer function constrained by the side‑bands. 
  \item Categories with the highest BDT response achieve $S/B\approx1.7\%$ (Cat. 4) inside a $\pm\,\sigma_{\rm core}$ window around 125 GeV, demonstrating the power of quark–gluon likelihood and soft‑activity variables introduced for this data-set.
  \item Systematics dominate over the limited luminosity: JES $6$–$10\%$, JER upto $10\%$; trigger $1$–$6\%$, $b$‑tag $3$–$9\%$; a $30\%$ prior on the $Z+$\emph{jets} yield, which is softened by the likelihood fit; functional choices for the multijet template remain the single largest contributor.
\end{enumerate}

\bigskip
\noindent
\textbf{CMS 13 TeV} (See Fig.~\ref{fig:thr-fou}):\\
Eighteen categories (\emph{Tight/Loose}, 2016/2018) are combined: The first eight panels show the $m_{b\bar b}$ distributions; the bottom-right panel displays the $S/(S\!+\!B)$‑weighted sum of all categories, where $S$ combines the \emph{VBF} and the \emph{ggF}.
\begin{enumerate}
  \item Constraining $ggF$ to the SM expectation, using $\mathcal{L}=36.3+54.4\;\mathrm{fb}^{-1}$ of Data, CMS measures $\mu_{\mathrm{VBF}} = 1.01^{+0.55}_{-0.46}$ with an observed significance of $2.4\,\sigma$ (2.7 $\sigma$ expected).
  \item For the mass spectra (first 8 panels), the consistency parameter for the analytic function ($\chi^{2}$) over the \emph{number of degrees of freedom (ndf)} is presented, together with the $p$-value that gives the probability (i.e., $p$) of observing the $\chi^{2}$ value as higher than actual. 
  \item The weighted spectrum (in the \emph{Data-QCD bkg.} plots) exhibits a clear $Z\!\to b\bar b$ (red) peak and a modest excess around 125 GeV (green) whose shape matches the calibrated \emph{Crystal‑Ball} signal model.
  \item Although still limited by statistics, several systematic terms are close in size: JES $7.7\%$ and JER $1.5\%$ impact on $\mu$; $6.7\%$ on the signal yield is from Trigger efficiency; $b$‑tagging calibration takes roughly $3\%$ after applying the DeepJet scale factors; $Z+$\emph{jets} normalisation takes $5$–$15\%$, treated as free nuisance parameters and constrained \emph{in situ} by the designated control categories; and the non‑resonant multijet background shape with the polynomial form is floated in the simultaneous fit (bias tests limit its effect to $\lesssim10\%$ of the statistical uncertainty, so it remains sub‑dominant). Residual theory uncertainties, such as those that depend on PDF, scale, and PS variations, each contribute below the $5\%$ level after profiling.
\end{enumerate}

\subsubsection*{Global picture and dominant uncertainties}
Across the four published \emph{VBF} $H\!\to\! b\bar b$ analyses,
all extracted signal‑strength modifiers are compatible with the SM
expectation $\mu=1$ within roughly $2\,\sigma$.  
The best measurement reaches a relative uncertainty of about $50\%$.

\begin{itemize}
  \item \textbf{Statistical component}: still the single largest contributor
        in every analysis.
  \item \textbf{Leading experimental systematics}: JES/JER, trigger efficiency, $b$‑tag calibration, and
        (in the CMS 13 TeV result) $b$‑jet energy regression and smearing.
  \item \textbf{Non‑resonant multijet background}: controlled either by
        profiling polynomial shape parameters inside the likelihood fit (in CMS)
        or by functional variations and side‑band statistics (in ATLAS); it can
        dominate when statistics are limited.
  \item \textbf{$Z+$\emph{jets} normalisation}: assigned sizeable priors
        (up to $25$--$50\%$) when left unconstrained, but reduced to
        $\lesssim15\%$ where the fit constrains it via CRs.
  \item \textbf{Signal‑theory uncertainties}: PDF, scale, and PS
        variations each impact $\mu$ at only the few‑percent level and are
        therefore sub‑dominant.
\end{itemize}

In summary, the $m_{b\bar b}$ spectra validate the data‑driven continuum description and confirm that the Run~1 sensitivity was statistics‑limited. The Run~2 samples at 13 TeV deliver a factor\,$\gtrsim$\,2 improvement in precision and start to probe the SM \emph{VBF} signal. No single measurement is yet conclusive, but taken together the results constitute the strongest direct constraints on $H\!\to\!b\bar b$ in the \emph{VBF} topology to date.

\subsection{Outlook}
\label{subsec:outlook}

With current $\mu_{\text{VBF}}$ precisions ($\sim 50\%$ for 13~TeV published analyses) remaining statistics-limited, Run~3 data-sets are expected to reduce statistical components while leaving certain systematics (JES/JER, $b$-tag calibration, $Z+$\emph{jets} modelling) as emerging limitations if not harmonised across \emph{VBF}, \emph{VH}, \emph{ggH}, and $t\bar t H$ channels. Multi-class categorisation (in CMS) and auxiliary channels (in ATLAS \emph{photon} channel) demonstrate routes to improved constraint of resonant and continuum backgrounds and may be extended or combined. Continued refinement of $Z+$\emph{jets} and non-resonant shape modelling, plus unified treatment of PS and \emph{Underlying Event (UE)} uncertainties, can be essential for preventing systematics from saturating gains in luminosity. 

Projected Run~3 improvements aim toward $\sim 30\%$ precision on $\mu_{\text{VBF}}$ and a correspondingly tighter constraint on $\kappa_b$ once combined with other production modes; present analyses already show that theory uncertainties on the \emph{VBF} signal are subdominant to experimental and statistical terms. Achieving these targets will require sustained optimisation of jet calibration, $b$-tagging, regression, and multi-class discrimination strategies, along with further decorrelation of production mode systematics in global fits (leveraging the $Z$- and $ggH$-control slices).

\section{Conclusion}



\emph{VBF} production followed by \(H\!\to b\bar{b}\) decay offers an electroweakly clean but statistics-limited probe of the bottom-quark Yukawa coupling. The present review collects the most recently published \emph{VBF} measurements performed by ATLAS and CMS at \(\sqrt{s}=8\) and \(13\;\text{TeV}\), arranging the
baseline selections, \emph{MVA} event categorisation, invariant-mass fits, and systematic treatments in a uniform format.  
This side-by-side presentation highlights the rapid methodological progress with the latest results.

Across all four analyses the signal is extracted from a simultaneous, binned maximum‑likelihood fit to the di‑$b$‑jet invariant mass, $m_{b\bar b}$; multivariate discriminants serve only to define orthogonal categories with differing $S/B$.  The smooth multijet continuum is described by data‑driven functions constrained in side‑bands, while the $Z\!\to b\bar b$ peak provides an internal calibration of mass resolution and overall modelling.  This strategy proves stable: in every data-set the post‑fit spectra reproduce the $Z$ resonance and show no spurious structures away from it.

Common features emerge.  First, percent‑level $S/B$ in the best categories underscores both the difficulty of the measurement and the effectiveness of quark–gluon separation, soft‑activity vetoes, and refined jet calibration.  Second, statistical errors still dominate the total uncertainty in all four results; systematic impacts (JES/JER, $b$‑tagging, $Z+$\emph{jets} modelling, PS variations) are subleading but non‑negligible.  Third, auxiliary or control categories (e.g.\ the ATLAS \emph{photon} channel, CMS \emph{Loose} $Z$ regions) tighten constraints on resonant and continuum components even when no visible Higgs peak is present.

The separate \emph{VBF} analyses at 13 TeV attain observed significances of \(1.9\sigma\)–\(2.4\sigma\) and yield cross-section central values that agree with the NNLO+NNLL SM prediction to within approximately \(20\,\%\). Statistical uncertainty is still dominant; the leading experimental systematics from the jet-energy calibration and heavy-flavour tagging are now
sub-dominant; this also applies for most of the theory uncertainties on the signal.

Taken together, these measurements deliver the most stringent direct constraints to date on $H\!\to b\bar b$ in the \emph{VBF} topology: Run~2 analyses achieve about a factor‑two improvement over Run~1, with ATLAS showing an upward fluctuation and CMS consistent with the SM within uncertainties.  The collective picture is one of methodological maturity, with efficient background control and stable mass fits, and steadily increasing precision, positioning the programme to pin down $\mu_{\text{VBF}}$ and, by extension, $\kappa_b$, with substantially reduced uncertainties as larger data-sets are incorporated.

The Yukawa interactions of the Higgs boson with the charged third-generation fermions (i.e., $t$, $b$, $\tau$) have each been directly established through observations of $t\bar{t}H$ / $tH$ production (top Yukawa), $H\to b\bar{b}$ decays in associated and \emph{VBF} production, and $H\to \tau^+\tau^-$ decays. However, the \emph{relative} (fractional) uncertainty on the bottom-quark Yukawa coupling $y_b$ (coupling modifier $\kappa_b$) remains larger than those on $\kappa_t$ and $\kappa_\tau$. This reduced precision principally reflects the difficulty of isolating the $H\to b\bar{b}$ signal from substantial QCD multijet and $V+$\emph{jets} backgrounds, necessitating elaborate multivariate analyses and finely optimized event categorizations. The extraction is further limited by experimental systematics (notably $b$-tagging efficiencies and mistag rates, JES, and JER) and by theoretical uncertainties affecting both signal and background modeling. In contrast, $\kappa_\tau$ benefits from cleaner leptonic final states with more favorable signal-to-background ratios, while $\kappa_t$ is concurrently constrained by multiple production modes ($t\bar{t}H$, $tH$) and its loop contributions to gluon-fusion and diphoton processes. Consequently, sharpening the precision on $\kappa_b$ remains a central objective of Run~3 and the HL-LHC physics programme.

No peer-reviewed \emph{VBF} \(H\!\to b\bar{b}\) result based on the ongoing
Run~3 data-set is yet available. Extrapolation of the Run~2 analysis framework to the expected
\(\mathcal{L}\simeq\!300\;\text{fb}^{-1}\) per experiment indicates that a \(30\,\%\) relative precision on \(\mu_{\mathrm{VBF}}\) is attainable, with a combined significance that could surpass \(3\sigma\). These gains derive almost entirely from larger Data samples, complemented by incremental improvements in object reconstruction and by
the retention of multivariate classifiers under Run~3 conditions.

Furthermore, with \(\mathcal{L}\simeq 3\;\text{ab}^{-1}\) of Data foreseen in the high-luminosity phase, the uncertainty on the \emph{VBF} signal strength can projectively fall to the \(15\,\%\) level. Such precision will enable stringent fits in both the SM Effective Field Theory and the Higgs (Electroweak) Effective Field Theory (i.e., SMEFT and HEEFT/HEFT), making the analyses sensitive to few‑percent modifications of the bottom Yukawa coupling. It will also allow a stand‑alone \(5\sigma\) observation of \emph{VBF} \(H\!\to b\bar{b}\); although, \(\kappa_{b}\) itself can still be dominated by the combination with \emph{VH} and \(t\bar t H\) channels. The channel will therefore play a decisive role in global Higgs‑coupling combinations and in testing potential deviations suggested by other production modes.

In summary, \emph{VBF} \(H\!\to b\bar{b}\) has evolved from an exploratory
search at 8 TeV into a precise measurement at 13 TeV.
The consolidated overview provided here establishes a reference
baseline against which upcoming Run~3 results can be assessed and
future refinements benchmarked, thereby closing the loop on the goals
outlined in the abstract and emphasising the continued importance of
this channel in the Higgs-coupling programme.

\pagebreak

\bibliographystyle{aip}
\bibliography{references}

\section*{Acknowledgements}
The authors received no financial supports or grants for this article.

\section*{Data Availability Statement}
No data is generated in this article.\\

\pagebreak

\section{Tables and Figures}

\subsection{The Properties of Higgs}


\begin{table}[h]
  \centering
  \begin{tabular}{|l|c|c|}
    \hline
    \textbf{Property} & \textbf{The SM prediction} & \textbf{Measurement} \\ \hline
    Mass \(m_{H}\) & \(\sim125\;\text{GeV}\) &
      \(125.20\pm0.11\;\text{GeV}\) \\
    Spin–parity \(J^{P}\) & \(0^{+}\) & \(0^{+}\) \\
    Full width \(\Gamma_{H}\) & \(4.07\;\text{MeV}\) &
      \(3.7^{+1.9}_{-1.4}\;\text{MeV}\) \\
    Coupling modifier \(\kappa_{F}\) (fermions) & 1 &
      \(0.94\pm0.05\) \\
    Coupling modifier \(\kappa_{V}\) (vector bosons) & 1 &
      \(1.023\pm0.026\) \\
    Coupling modifier \(\kappa_{b}\) (bottom quark) &  1  &  \( 0.99\pm0.3\)\\ 
    $H$ signal strength ($b\bar{b}$-channel) &  1  &  $0.99\pm 0.12$ \\
    Production cross-section (13 TeV $p\bar{p}$)  &  54-55  &  $54\pm 2.6$ pb \\ \hline
  \end{tabular}
  \caption{Key Higgs–boson properties and coupling modifiers. The shown measurements, from 13 TeV analyses, are the closest that match the SM predictions; for more information on the measurements see \cite{pdgHiggs23, ParticleDataGroup:2024cfk,HXSWG:2017cgw,CMS:2022bmu,ATLAS:2022vkf}}
  \label{tab:1}
\end{table}


\subsection{Analysis Selections}


\begin{table}[htbp]
  \centering
  \begin{tabular}{|p{4.1cm}|p{10.4cm}|}
    \hline
    \textbf{Stage} & \textbf{Requirement} \\ \hline
    L1 multi-jet          & $\ge 4$ jets with $p_{\mathrm{T}} > 15$ GeV \\ 
    L1 \emph{VBF}-Higgs alt.     & $(3$ jets $+$ 1  fwd jet ($\mid\eta\mid>3.2$)) or $(2$ fwd jets) each $p_{\mathrm{T}} > 15$ GeV \\ 
    HLT                   & Same 4 jets with $p_{\mathrm{T}} > 35$ GeV; $\ge 2$ HLT $b$-tagged jets  \\ 
    Offline jets          & Exactly 4 jets, $p_{\mathrm{T}} > 50$ GeV, $|\eta| < 4.5$; outer pair = \emph{VBF} jets, \qquad      inner pair = Higgs ($b\bar{b}$-decay) jets \\ 
    Pre-selection $b$-tagging           & 2 Higgs jets within tracker acceptance ($|\eta| < 2.5$) that match the HLT $b$-jets (passing $70\%$ efficiency \emph{Working Point (WP)}) \\ 
    Kinematics            & $p_{\mathrm{T}}^{b\bar{b}} > 100$ GeV to flatten multijet $m_{b\bar{b}}$ \\ \hline
  \end{tabular}
  \caption{ATLAS 8 TeV \emph{VBF} $H\!\to\!b\bar b$ – full trigger and baseline selection applied to Data and MC \cite{ATLAS:2016mzy}. \emph{Level 1 (L1)} trigger, \emph{High Level Trigger (HLT)}, as well other analysis selections are listed.}
  \label{tab:2A8}
\end{table}



\begin{table}[htbp]
  \centering
  \begin{tabular}{|p{2.4cm}|p{4.9cm}|p{6.7cm}|}
    \hline
    \textbf{Channel} & \textbf{Trigger (L1 / HLT)} & \textbf{Offline baseline} \\ \hline
    \emph{Two-central} &
    \textbf{L1:} $\ge 2$ central ($\mid\eta\mid<2.8$) jets ($E_{\mathrm{T}}>$ 40, 25 GeV) + \qquad\qquad fwd ($3.2<\mid\eta\mid<4.4$) jet \qquad\qquad  ($>$ 20 GeV);  
    \textbf{HLT:} $\ge 2$ $b$-jets \qquad\qquad
    ($>$ 80, 60 GeV; 70$\%$, 85$\%$ WP) + fwd jet ($>$ 45 GeV) &
    $\geq 4$ jets ($p_{\mathrm{T}}>$20 GeV, $\mid\eta\mid<4.4$) with \qquad\qquad
    $\ge 2$ $b$-jets ($p_{\mathrm{T}}>$ 95, 70 GeV, $|\eta|<$2.4, 2.5; $70\%$, $85\%$ WP); $\geq$1 fwd jet ($p_{\mathrm{T}}>60$ GeV); $\geq$1 jet ($p_{\mathrm{T}}>20$ GeV, $\mid\eta\mid < 4.4$);  
    $p_{\mathrm{T}}^{b\bar{b}}>160$ GeV \\ \hline
    \emph{Four-central} &
    \textbf{L1:} $\ge 4$ jets ($E_{\mathrm{T}}>$ 15 GeV); 
    \textbf{HLT:} $\ge 2$ $b$-jets ($>$ 45, 35 GeV; $70\%$, $60\%$ WP) &
    $\ge 2$ $b$-jets ($p_{\mathrm{T}}>55$ GeV, $\mid\eta\mid<$ 2.5; $70\%$ WP) + $\ge 2$ central jets( $p_{\mathrm{T}}>55$ GeV);  
    fwd jet veto if $p_{\mathrm{T}}>60$ GeV;  
    $p_{\mathrm{T}}^{b\bar{b}}>150$ GeV \\ \hline
    Photon + \emph{VBF} &
    \textbf{L1:} $\gamma$ ($E_{\mathrm{T}}>22$ GeV);  
    \textbf{HLT:} $\gamma$ ($>25$ GeV) + $\ge 4$ jets \qquad\qquad or $\ge 3$ jets and 1 $b$-jet at $77\%$ WP ($>35$ GeV and $\mid\eta\mid<4.9$); $m_{jj}>700$ GeV for $\geq$1 dijet pair & 
    $\geq 1 \gamma$ ($E_{\mathrm{T}}>30$ GeV, $\mid\eta\mid 1.37$ or 1.52 \qquad\qquad $<\mid\eta\mid<2.37$) +  
    $\ge 2$ $b$-jets ($p_{\mathrm{T}}>40$ GeV, $\mid\eta\mid <2.5$; $77\%$ WP) + $\ge 2$ jets ($p_{\mathrm{T}}>40$ GeV, $\mid\eta\mid <4.4$); $m_{jj}>800$ GeV;  
    $p_{\mathrm{T}}^{b\bar{b}}>80$ GeV \\ \hline
  \end{tabular}
  \caption{ATLAS 13 TeV \emph{VBF} $H\!\to\!b\bar b$ – trigger and baseline cuts for all three channels \cite{ATLAS:2018jvf}. All energy cuts are on the transverse energy scale ($E_{\mathrm{T}}$) unless otherwise written (i.e., on $p_{\mathrm{T}}$). \emph{Level 1 (L1)} trigger, \emph{High Level Trigger (HLT)}, as well other analysis selections are listed.}
  \label{tab:2A13}
\end{table}



\begin{table}[htbp]
  \centering
  \begin{tabular}{|p{2.1cm}|p{6.2cm}|p{5.8cm}|}
    \hline
    \textbf{Channel} & \textbf{Trigger (L1 / HLT)} & \textbf{Offline baseline} \\ \hline
    
    Dedicated $qqH\rightarrow qqb\bar{b}$ \emph{VBF} (Set A) &
    \textbf{L1:} $\ge 3$ jets with thresholds $p_{\mathrm{T}} \!=\!64$–68,\;44–48,\;24–32 GeV;  
    exactly one of the two lead jets may be fwd ($2.6<|\eta|\le5.2$), others central ($|\eta|\le2.6$).  
    \textbf{HLT:} $\ge4$ jets with $p_{\mathrm{T}}>$ 75–82,\;55–65,\;35–48,\;20–35 GeV,  
    $\ge1$ online $b$-jet (TCHE or CSV);  
    \emph{VBF}‑tag jet pair either with the smallest $b$‑tag values or the largest $|\Delta\eta_{qq}|$;  
    require $|\Delta\eta_{qq}|>2.2$–2.5  and $m_{qq}>200$–240 GeV  &
    
    $\ge4$ PF jets with $p_{\mathrm{T}}>80,\;70,\;50,\;40$ GeV and $|\eta|<4.5$;  
    $\ge2$ jets CSV loose (CSVL);  
    \emph{VBF} topology: $m_{qq}>250$ GeV,\; $|\Delta\eta_{qq}|>2.5$;  
    two $b$‑jets with $\Delta\phi_{b\bar{b}}<2.0$ rad  \\ \hline
    
    General‑ purpose \emph{VBF} (Set B) &
    \textbf{L1:} scalar $H_{\mathrm{T}}$ of calorimeter jets $>175$ – 200 GeV.  
    \textbf{HLT:} $\ge2$ CaloJets ($p_{T}>35$ GeV);  
    pick the opposite‑hemisphere jet pair with largest $m_{jj}$  
    ($m^{\text{trig}}_{jj}>700$ GeV,\; $|\Delta\eta^{\text{trig}}_{jj}|>3.5$) &
    
    (Events fail Set A)  
    $\ge4$ PF jets with $p_{\mathrm{T}}>30$ GeV,\; $|\eta|<4.5$;  
    $H_{T}^{j1.j2}>160$ GeV;  
    $b$‑tag: $\ge1$ CSV medium (CSVM) jet with $\ge1$ CSVL jet;  
    \emph{VBF} topology: $m_{qq},\,m^{\text{trig}}_{jj}>700$ GeV and $|\Delta\eta_{qq}|,\,|\Delta\eta^{\text{trig}}_{jj}|>3.5$;  
    $\Delta\phi_{b\bar{b}}<2.0$ rad  \\ \hline
  \end{tabular}
  \caption{CMS 8 TeV \emph{VBF} $H\!\to\!b\bar b$ – trigger paths and baseline cuts for the two data-sets used in the analysis \cite{CMS:2015two}. Jet $p_{T}$ values are given at the electromagnetic or particle‑flow (PF) scale as appropriate. Online thresholds varied slightly with instantaneous luminosity, where a dedicated $qqH\rightarrow qqb\bar{b}$ establishes either the \emph{track counting high-efficiency (TCHE)} or the \emph{combined secondary vertex (CSV)} algorithms for HLT. \emph{Level 1 (L1)} trigger, \emph{High Level Trigger (HLT)}, as well other analysis selections are listed.}
  \label{tab:CMS8TeV_VBF}
\end{table}



\begin{table}[htbp]
  \centering
  \begin{tabular}{|p{1.0cm}|p{1.1cm}|p{5.2cm}|p{5.9cm}|}
    \hline
    \textbf{Year} & \textbf{Path} & \textbf{HLT jets / $b$-tags} & \textbf{Offline baseline} \\ \hline
    2016 & 
        \emph{Tight} & 
    Events with 4 jets ($p_{\mathrm{T}}>$ 92, 76, 64, 15 GeV); $\ge\!1$ online $b$‑jet; \emph{VBF} topology ($m_{jj}\!>\!500$ GeV; $\Delta\eta_{jj}\!>\!4.1$; $\Delta\phi_{b\bar{b}}\!<\!1.6$ rad) &
    Events with 4 jets ($p_{\mathrm{T}}>$ 95, 80, 65, 30 GeV), \emph{VBF} adjusted from HLT ($\Delta\eta_{jj}\!>\!4.2$); offline standard (jets with $\eta<4.7$;
    $\geq$2 DEEPJET $b$-jets with $\eta<2.4$; veto leptons)\\ \hline
    2016     & \emph{Loose} & 
    Same 4 jets as \emph{Tight}; $\ge\!2$ online $b$‑jet; \emph{VBF} topology ($m_{jj}\!>\!240$ GeV; $\Delta\eta_{jj}\!>\!2.3$; $\Delta\phi_{b\bar{b}}\!<\!2.1$ rad) &
    Same 4 jets as \emph{Tight}; \emph{VBF Loose} offline topology ($m_{jj}\!>\!250$ GeV; $\Delta\eta_{jj}\!>\!2.5$; $\Delta\phi_{b\bar{b}}\!<\!2.1$ rad); offline standard  \\ \hline
    2018 &     \emph{Tight} & 
    Events with 4 jets ($p_{\mathrm{T}}>$ 105, 88, 76, 15 GeV); $\ge\!1$ online $b$‑jet; \emph{VBF} topology ($m_{jj}\!>\!460$ GeV; $\Delta\eta_{jj}\!>\!3.5$; $\Delta\phi_{b\bar{b}}\!<\!1.9$ rad) &
    Events with 4 jets ($p_{\mathrm{T}}>$ 110, 90, 80, 30 GeV), \emph{VBF} topology ($m_{jj}\!>\!500$ GeV; $\Delta\eta_{jj}\!>\!3.8$; $\Delta\phi_{b\bar{b}}\!<\!1.6$ rad); offline standard \\ \hline
    2018     & \emph{Loose} & 
    Same 4 jets as \emph{Tight}; $\ge\!2$ online $b$‑jet; \emph{VBF} topology ($m_{jj}\!>\!200$ GeV; $\Delta\eta_{jj}\!>\!1.5$; $\Delta\phi_{b\bar{b}}\!<\!2.8$ rad) &
    Same 4 jets as \emph{Tight}; \emph{VBF Loose} offline topology; offline standard\\ \hline
  \end{tabular}
  \caption{CMS 13 TeV \emph{VBF} $H\!\to\!b\bar b$ – trigger and baseline cuts for \emph{Tight} and \emph{Loose} paths in 2016 and 2018 \cite{CMS:2023tfj}.  \emph{High Level Trigger (HLT)} as well other analysis selections are listed.}
  \label{tab:2C13}
\end{table}


\pagebreak

\subsection{BDT Strategies}



\begin{table}[htbp]
  \centering
  \begin{tabular}{|l|c|c|c|c|c|}
    \hline
    \textbf{Cat} & \textbf{BDT window $w$} & \textbf{Data events} & \textbf{\emph{VBF}}  & \textbf{\emph{ggF}}  & \textbf{$Z+$\emph{jets}} \\ \hline
    Pre‑selection & —                       & 554\,302 & 130 & 94   & 3\,700 \\ \hdashline
    I         & $-0.08 < w \leq 0.01$      & 176\,073 & 39  & 31   & 1\,100 \\ \hdashline
    II        & $0.01 < w \leq 0.06$       & 46\,912  & 33  & 8.5  & 350 \\ \hdashline
    III       & $0.06 < w \leq 0.09$       & 15\,015  & 23  & 3.8  & 97 \\ \hdashline
    IV        & $w > 0.09$              & 6\,493   & 19  & 1.6  & 49 \\ \hline
  \end{tabular}
  \caption{ATLAS 8 TeV \emph{VBF} \(H\!\to\!b\bar b\) – BDT \emph{Category (Cat.)} windows and event yields for $70 < m_{b\bar{b}} < 300$ GeV; events are counted after passing the Trigger and offline selections \cite{ATLAS:2016mzy}.}
  \label{tab:3A8}
\end{table}


\begin{table}[htbp]
  \centering
  \begin{tabular}{|l|c|c|r|r|r|r|}
    \hline
    \textbf{Channel} & \textbf{SR} & \textbf{BDT window $w$} & \textbf{Data} & \textbf{Higgs boson} & \textbf{$Z$+jets} & \textbf{Non‑res bkg} \\ \hline
    \emph{Two‑central} & I   & $w \ge -0.006$          & 35\,496 & 340 & 470 & 34\,620 \\ \hdashline
                & II  & $w < -0.006$            & 95\,802 & 165 & 230 & 95\,620 \\ \hline
    \emph{Four‑central} & I   & $w > 0.033$             & 13\,139 & 167 & 22  & 12\,870 \\ \hdashline
                & II  & $0.026 < w \le 0.033$   & 19\,611 & 101 & 197 & 19\,340 \\ \hdashline
                & III & $0.015 < w \le 0.026$   & 60\,314 & 183 & 720 & 59\,340 \\ \hdashline
                & IV  & $0.002 < w \le 0.015$   & 148\,413 & 304 & 1\,260 & 146\,930 \\ \hline
    Photon+\emph{VBF}  & I   & $w > 0.30$              & 162   & 21.1 & 5.8 & 140.4 \\ \hdashline
                & II  & $-0.05 \le w \le 0.30$  & 565   & 20.1 & 1.1 & 518 \\ \hdashline
                & III & $w < -0.05$             & 1\,270 & 10.6 & 9.8 & 1\,296 \\ \hline
  \end{tabular}
  \caption{ATLAS 13 TeV \emph{VBF} \(H\!\to\!b\bar b\) – BDT \emph{Signal Region (SR)} windows and event yields for $100 < m_{b\bar{b}} < 140$ GeV; events are counted after passing the Trigger and offline selections
  \cite{ATLAS:2018jvf}. All \emph{Higgs boson} (\emph{VBF, ggF, VH,} $t\bar{t}H$) events are included, next to $Z+$\emph{jets} and \emph{non-resonant backgrounds}. }
  \label{tab:3A13}
\end{table}




\begin{table}[htbp]
  \centering
  \begin{tabular}{|l|c|c|r|r|r|r|r|}
    \hline
    \textbf{Selection} & \textbf{Cat.} & \textbf{BDT window $D$} &
    \textbf{Data} &
    \textbf{\emph{VBF}} & \textbf{ggF} & \textbf{$Z$+jets} &
    \textbf{$W$+jets \& top bkg} \\ \hline
    Set A & 1 & $-0.6 < D \leq 0.0$   & 546\,121 & 53 & 53 & 2\,038 & 4\,060 \\ \hdashline
          & 2 & $0.0 < D \leq 0.7$    & 321\,039 & 140 & 51 & 1\,584 & 1\,607 \\ \hdashline
          & 3 & $0.7 < D \leq 0.84$   & 32\,740  & 58 & 8  & 198   & 113 \\ \hdashline
          & 4 & $0.84 < D \leq 1.0$   & 10\,874  & 57 & 5  & 71    & 39 \\ \hline
    Set B & 5 & $-0.1 < D \leq 0.4$   & 203\,865 & 33 & 9  & 435   & 761 \\ \hdashline
          & 6 & $0.4 < D \leq 0.8$    & 108\,279 & 57 & 10 & 280   & 420 \\ \hdashline
          & 7 & $0.8 < D \leq 1.0$    & 15\,151  & 31 & 2  & 45    & 68 \\ \hline
  \end{tabular}
  \caption{CMS 8 TeV \emph{VBF} \(H\!\to\!b\bar b\) – BDT \emph{Category (Cat.)} windows and event yields for $80 < m_{b\bar{b}} < 200$ GeV; events are counted after passing the Trigger and offline selections
  \cite{CMS:2015two}.}
  \label{tab:3C8}
\end{table}




\begin{table}[htbp]
  \centering\small
  \begin{tabular}{|l|c|p{4.4cm}|r|r|r|r|r|}
    \hline
    \textbf{Sample} & \textbf{Cat.} & \textbf{BDT window} &
    \textbf{Data} & \textbf{\emph{VBF}} & \textbf{ggF} & \textbf{$Z$+jets} &
    \textbf{$W$+jets \& top bkg} \\ \hline
    2016 \emph{Loose} & G1 & $0.50 < D_{\mathrm{ggH}} < 0.55$ &
      41\,432 & 4.5 & 27.5 & 275 & 127.3 \\ \hdashline
               & G2 & $D_{\mathrm{ggH}} > 0.55$ &
      58\,895 & 6.1 & 51.6 & 407 & 135.5 \\ \hdashline
               & V1 & $D_{\mathrm{ggH}}<0.50,\; 0.80< D_{\mathrm{VBF}}<0.85$ &
       4\,330 & 19.9 & 2.7  & 45  & 12.5 \\ \hdashline
               & V2 & $D_{\mathrm{ggH}}<0.50,\; D_{\mathrm{VBF}}>0.85$ &
       1\,901 & 17.4 & 1.7  & 31  & 5.2 \\ \hdashline
               & Z1 & $D_{\mathrm{ggH}}<0.50,\; D_{\mathrm{VBF}}<0.80,$\newline $0.60<D_Z<0.75$ &
      78\,850 & 9.6 & 34.8 & 1\,150 & 267.1 \\ \hdashline
               & Z2 & $D_{\mathrm{ggH}}<0.50,\; D_{\mathrm{VBF}}<0.80,$\newline $D_Z>0.75$ &
      29\,992 & 3.1 & 14.1 & 650 & 234.0 \\ \hline
    2016 \emph{Tight} & 1 & $0.25 < D < 0.50$ &
      29\,864 & 92.7 & 15.6 & 161 & 45.3 \\ \hdashline
               & 2 & $0.50 < D < 0.75$ &
      21\,831 & 136.2 & 13.9 & 151 & 26.7 \\ \hdashline
               & 3 & $D > 0.75$ &
       7\,231 & 117.3 & 6.2 & 75 & 8.6 \\ \hline
    2018 \emph{Loose} & G1 & $0.55 < D_{\mathrm{ggH}} < 0.60$ &
      17\,296 & 2.4 & 13.5 & 137 & 34.4 \\ \hdashline
               & G2 & $D_{\mathrm{ggH}} > 0.60$ &
      24\,882 & 2.9 & 24.3 & 180 & 35.2 \\ \hdashline
               & V1 & $D_{\mathrm{ggH}}<0.55,\; 0.50< D_{\mathrm{VBF}}<0.55$ &
       1\,914 & 6.4 & 1.0 & 22 & 4.9 \\ \hdashline
               & V2 & $D_{\mathrm{ggH}}<0.55,\; D_{\mathrm{VBF}}>0.55$ &
       2\,453 & 11.0 & 1.7 & 25 & 5.0 \\ \hdashline
               & Z1 & $D_{\mathrm{ggH}}<0.55,\; D_{\mathrm{VBF}}<0.50,$\newline $0.60<D_Z<0.70$ &
      24\,559 & 7.0 & 10.8 & 506 & 78.5 \\ \hdashline
               & Z2 & $D_{\mathrm{ggH}}<0.55,\; D_{\mathrm{VBF}}<0.50,$\newline $D_Z>0.70$ &
      14\,530 & 4.0 & 7.1 & 445 & 123.0 \\ \hline
    2018 \emph{Tight} & 1 & $0.25 < D < 0.50$ &
      29\,261 & 89.5 & 18.0 & 190 & 57.3 \\ \hdashline
               & 2 & $0.50 < D < 0.75$ &
      23\,392 & 139.8 & 17.9 & 202 & 38.8 \\ \hdashline
               & 3 & $D > 0.75$ &
       8\,202 & 134.5 & 8.7 & 104 & 11.0 \\ \hline
  \end{tabular}
  \caption{CMS 13 TeV \emph{VBF} \(H\!\to\!b\bar b\) – BDT \emph{Category (Cat.)} windows and event yields for combined regions of $80 < m_{b\bar{b}} < 104$ GeV and $146 < m_{b\bar{b}} < 200$ GeV; events are counted after passing the Trigger and offline selections
  \cite{CMS:2023tfj}.}
  \label{tab:3C13}
\end{table}


\pagebreak

\subsection{Results and Uncertainties}
\label{subsec:fiv}


\begin{table}[htbp]
  \centering
  \renewcommand{\arraystretch}{1.15}
  \begin{tabular}{|l|c|l|c|c|c|c|}
    \hline
    \textbf{Experiment} & $\sqrt{s}$ & \textbf{Fit variant} &
    $\boldsymbol{\mu}$ &
    $\mathbf{UL_{95}^{obs}}$ &
    $\mathbf{UL_{95}^{exp}}$ &
    \textbf{Significances $[\sigma]$}  \\ 
      & [TeV] & \textbf{/ channel} & (cut-based) & & \emph{(with SM)} & 
      \\ 
    \hline\hline
    ATLAS & 8  & \emph{VBF}                  & $-0.8 \pm 2.3$         & 4.4 & 5.4 & —     \\
      &    &  & ($-5.2^{+4.6}_{-4.4}$) &  & \emph{(5.7)} &     \\\hline

    ATLAS & 13 & \emph{VBF}                          & $3.0^{+1.7}_{-1.6}$    & 5.9 & 3.0 &   Obs.: 1.9, Exp.: 0.7 \\
           &    & Inclusive                          & $2.5^{+1.4}_{-1.3}$    & 4.8 & 2.5 &  Obs.: 1.9, Exp.: 0.8 \\ \hline

    CMS   & 8  & \emph{VBF}                            & $2.8^{+1.6}_{-1.4}$    & 5.5 & 2.5 & Obs.: 2.20, Exp.: 0.83  \\
       &    & Inclusive                 & $1.03^{+0.44}_{-0.42}$ & 1.77 & 0.78 & Obs.: 2.56, Exp.: 2.70  \\ \hline
       
    CMS   & 13 & \emph{VBF}               & $1.01^{+0.55}_{-0.46}$ & —   & —   & 
    Obs.: 2.4, Exp.: 2.7\\
           &    & Inclusive                & $0.99^{+0.48}_{-0.41}$ & —   & —   & Obs.: 2.6, Exp.: 2.9  \\ \hline
  \end{tabular}
  \caption{Best-fit signal strengths ($\mu$), observed and expected 
           $95\,\%$ Confidence Level \emph{Upper Limits}, ($UL_{95}$), on $\sigma\times\mathcal B$, and 
           \emph{observed/expected (obs/exp)} significances for \emph{VBF} $H\!\to\!b\bar b$ at 
           8 TeV and 13 TeV \cite{ATLAS:2016mzy,ATLAS:2018jvf,CMS:2015two,CMS:2023tfj}.  
           A dash (—) indicates the quantity was not quoted in the 
           corresponding publication.}
  \label{tab:6}
\end{table}


\begin{table}[htbp]
  \centering\small
  \renewcommand{\arraystretch}{1.15}
  \begin{tabular}{|p{2.7cm}|p{2.9cm}|p{2.9cm}|p{2.9cm}|p{2.9cm}|}
    \hline
    \multirow{2}{*}{\textbf{Source}} &
      \multicolumn{2}{c|}{\textbf{ATLAS}} &
      \multicolumn{2}{c|}{\textbf{CMS}} \\ \cline{2-5}
      & 8 TeV  & 13 TeV  & 8 TeV  & 13 TeV \\ \hline\hline
    \textbf{Data statistics}               & MVA: $\pm1.3$ (Cut-based: $\pm3.7$)       & \emph{VBF}: $\pm 0.9$ (incl.:$\pm 0.6$) & not separated$^{\dagger}$ & not separated$^{\dagger}$ \\ \hline\hline
    \textbf{\textit{Systematic uncertainties:}} &  \emph{MVA (Cut-based) values } & \emph{VBF (Incl.)}  & \textit{Percent value compared with signal ($\%$)} &  Percent value compared with signal ($\%$) \\ \hline
    \textbf{Theory}     & MC signal model.: $\pm 0.1$ ($\pm 1.3$), $Z$ yield: $+0.6/$ $-0.5$ ($\pm 1.4$)         & Higgs model.: $+0.2/$ $-0.1$ ($+0.3/$ $-0.1$)            & PS $\&$ UE: 2-7; PDF (global, categ.): 2.8, 1.5-3; Scale var. (global, categ.): 0.2, 1-5   & \emph{VBF} PS: 13.0, PS (final-state rad.): 5.6, Pileup: 2.3\\ \hline
    \textbf{Non‑res.\ bkg.\ model}         & Fnc. choice: $\pm 1.0$ ($\pm 1.0$), Sideband stat.:$\pm 1.7$ ($\pm 3.7$)    & $\pm 1.2$ ($\pm 1.0$)       & fitted (dominant)         & fitted \qquad\qquad\qquad\qquad (sub-dominant) \\ \hline
    \textbf{ $Z+$\emph{jets} normalisation}      & Percent ($\%$) uncertainty due to higher order QCD corrections to $Z\rightarrow b\bar{b}$: 25 (uncorrelated: 40-50)   & $\pm 0.5$ ($\pm 0.5$)           & $30$ (constraint)         & 5-15 (treated in‐fit) \\ \hline
    \textbf{\textit{Sys. unc. affecting the signal:}} & \textit{Impact on signal strength ($\%$)}   &   & \textit{Impact on signal strength ($\%$)} &  \textit{Impact on signal strength ($\%$)} \\ \hline
    \textbf{Jet‑energy scale / resolution} &  JES: 15, JER: 4   & JES $\&$ JER: $+0.4/$ $-0.2$ ($+0.3/$ $-0.2$)    & JES (acceptance, signal shape): 6-10, 2; JER (acc., sig.sh.): 1-4, 10 & JES: 7.7, JER: 1.5 \\ \hline
   \textbf{ Trigger /} $\boldsymbol{b}$\textbf{‑tag}  &  $b$jet - trig: 10-20, tag: 10  & $b$-tag $\&$ trig: $+0.2/$ $-0.1$ ($+0.2/$ $-0.1$)  & trig.:1-6, $b$-jet tag: 3-9  & trig eff.: 6.7,  
    $b$-jet regres. smear.: 3.3, $b$-tag eff.: 3.0, $b$-jet regres. scale: 2.0 \\ 
    \hline
    \textbf{Other}  &  Experimental Unc. - detector related: $+0.2/$ $-0.3$ ($+1.6/$ $-1.2$), MC stat.: $\pm 0.4$ ($\pm 0.1$)   &   &  Integ. lumi: 1.9-2.6, $H\rightarrow b\bar{b}$ branch. fract.: 2.4-4.3, q/g jet tag: 1-3 & \\ \hline
    \multicolumn{5}{l}{\footnotesize$^{\dagger}$ Only the total fitted uncertainty on $\mu$ is given.} \\
  \end{tabular}
  \caption{Leading contributions to the overall uncertainty for the four \emph{VBF} $H{\to}b\bar b$ searches \cite{ATLAS:2016mzy,ATLAS:2018jvf,CMS:2015two,CMS:2023tfj}. \emph{Parton Shower (PS)}, \emph{Underlying Event (UE)}, \emph{Jet Energy Scale/Resolution (JES/JER)}, \emph{trigger (trig)}, and other contributors to uncertainties are presented.}
  \label{tab:7}
\end{table}



\end{document}